\documentclass[lettersize,journal]{IEEEtran}
\usepackage{amsmath,amsfonts}
\usepackage{array}
\usepackage{textcomp}
\usepackage{stfloats}
\usepackage{url}
\usepackage{verbatim}
\usepackage{graphicx}
\usepackage{cite}
\usepackage{multirow} 
\usepackage{bm}

\usepackage{graphicx} 
\usepackage[linesnumbered,ruled,vlined]{algorithm2e}
\usepackage{amsmath} 
\usepackage{amssymb}

\usepackage{array}

\usepackage{enumerate}
\usepackage{enumitem}
\usepackage{makecell}
\usepackage[labelformat=empty]{subcaption}
\usepackage[belowskip=3pt, aboveskip=3pt]{caption}
\usepackage{booktabs}
\usepackage{xcolor}

\begin{document}

\title{Community Search in Time-dependent Road-social Attributed Networks}

\author{Li Ni, Hengkai Xu, Lin Mu, Yiwen Zhang*, and Wenjian Luo, ~\IEEEmembership{Senior Member,~IEEE}
\thanks{Li Ni, Hengkai Xu, Lin Mu, and Yiwen Zhang are with the School of Computer Science and Technology, Anhui University, Hefei, Anhui, 230601, China. 

Wenjian Luo is with the School of Computer Science and Technology,
Harbin Institute of Technology, Shenzhen, 518055, China.

Email: nili@ahu.edu.cn, xhk@stu.ahu.edu.cn, mulin@ahu.edu.cn, zhangyiwen@ahu.edu.cn, and luowenjian@hit.edu.cn. (Corresponding author: Yiwen Zhang)}
\thanks{}}

\markboth{}%
{}

\IEEEpubid{}

\maketitle

\begin{abstract}
Real-world networks often involve both keywords and locations, along with travel time variations between locations due to traffic conditions.
However, most existing cohesive subgraph-based community search studies utilize a single attribute, either keywords or locations, to identify communities.
They do not simultaneously consider both keywords and locations, which results in low semantic or spatial cohesiveness of the detected communities, and they fail to account for variations in travel time.
Additionally, these studies traverse the entire network to build efficient indexes, but the detected community only involves nodes around the query node, leading to the traversal of nodes that are not relevant to the community.
Therefore, we propose the problem of discovering semantic-spatial aware $k$-core, which refers to a $k$-core with high semantic and time-dependent spatial cohesiveness containing the query node. 
To address this problem, we propose an exact and a greedy algorithm, both of which gradually expand outward from the query node.
They are local methods that only access the local part of the attributed network near the query node rather than the entire network. 
Moreover, we design a method to calculate the semantic similarity between two keywords using large language models. This method alleviates the disadvantages of keyword-matching methods used in existing community search studies, such as mismatches caused by differently expressed synonyms and the presence of irrelevant words.
Experimental results show that the greedy algorithm outperforms baselines in terms of structural, semantic, and time-dependent spatial cohesiveness.

\end{abstract}


\begin{figure}[!t]
  \centering
  \includegraphics[width=\linewidth]{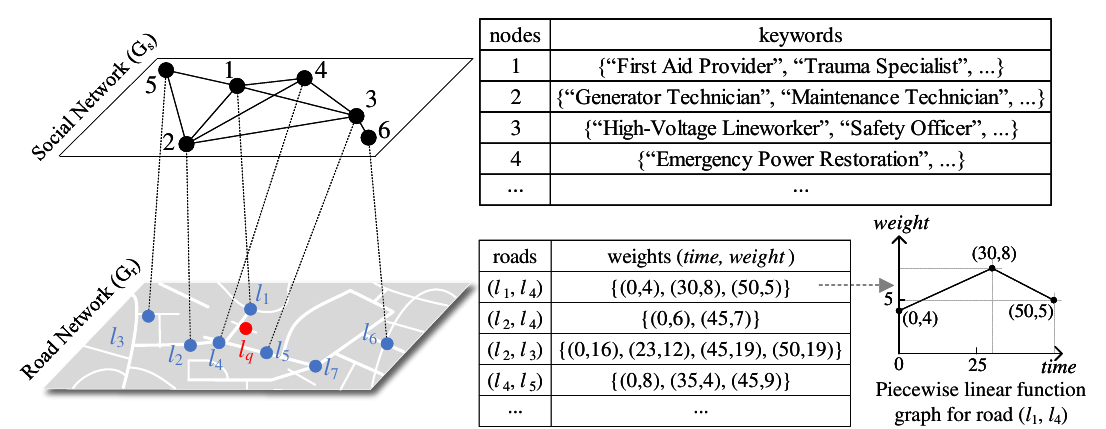}
  \caption{A toy example of time-dependent road-social attributed networks. Each node in the social network is associated with keywords and a location on the road network. The travel time (weight) of each road follows a piecewise linear function. For example, the weight $w_{l_{1}, l_{4}} = {(0, 4), (30, 8), (50, 5)}$ for the road $(l_{1}, l_{4})$ is represented by the linear function in the lower right part of the figure, where $(0, 4)$ indicates the travel time between $l_{1}$ and $l_{4}$ is 4 minutes at time 0.}
  \label{figure:toy}
\end{figure}

\section{Introduction}
With the growing prevalence of location-based services \cite{Chen_01, Fang_03}, road-social networks have emerged \cite{Wang_01, Guo_03, Qiyuan_01}. In these networks, users are associated with various attributes, such as keywords \cite{Cheng_01} and locations \cite{Zhou_01, XiongYPKF24, ChenWZHLZZW23}, and the travel times between locations in real road networks vary based on traffic conditions. These networks are referred to as time-dependent road-social attributed networks \cite{Wang_01}, as illustrated in Figure \ref{figure:toy}.

In this article, we study the problem of Semantic-Spatial Aware $k$-Core (SSAC) search. Specifically, given a time-dependent road-social attributed network, a set of keywords, a location, and a query node, our objective is to identify an SSAC containing the query node.
Essentially, the SSAC satisfies structural, semantic, and time-dependent spatial cohesiveness, meaning that nodes within the SSAC are tightly connected, have keywords semantically related to the query keywords, and reach the query location with short travel times.
Figure \ref{figure:toy} illustrates an SSAC with four users $\{1, 2, 3, 4\}$, who are densely interconnected, have keywords semantically related to “power maintenance”, and reach the query location $l_q$ with short travel times.
Such search capabilities enable the quick organization of impromptu offline activities. Potential applications include:
\begin{itemize}[leftmargin=*]
    \item Assemble the emergency power outage response team. Team members should meet the following requirements: (1) To ensure effective repairs, they preferably possess “power maintenance” skills. (2) To reduce coordination costs, they should have prior collaboration experience. (3) To minimize downtime, members should be stationed near the fault location to ensure arrival on site within a 10-minute response time.
    \item Detecting fake account clusters. To serve purposes such as marketing or public opinion manipulation, fraudulent groups often control numerous accounts to forward the same content, creating dense follow and forward relations. 
    Due to cost and technical constraints, these accounts are typically managed in bulk, resulting in the location of controlled accounts' IP addresses within specific regions.
    These fake accounts form communities characterized by frequent interactions, similar content, and geographic proximity.
    
\end{itemize} 

\textbf{Prior works.} Community search based on both structure and attributes, typically classified into keyword- or location-based community search. For keyword-based community search, researchers identify communities with both structural and keyword cohesiveness using shared keywords and scoring functions \cite{Fang_01, Huang_01}. Similarly, location-based community search focuses on communities with high structural and spatial cohesiveness \cite{Chen_01, Fang_03, Fang_04}. Researchers employ the minimum covering circle \cite{Fang_03} and spatial co-location \cite{Chen_01} to make communities with high spatial cohesiveness.
The above studies are effective in uncovering communities that exhibit keyword or spatial cohesiveness. 
However, they still suffer from several limitations in uncovering community structure in time-dependent road-social attributed networks, as discussed below.

\textbf{Limitations. }
1) The above studies utilize a single attribute, either keywords or locations (using Euclidean distance to estimate spatial proximity), neglecting the fact that users possess both keywords and locations, as well as variable travel times. Consequently, this leads to low semantic or spatial cohesiveness of the detected communities and the failure of these studies to adapt to variations in travel time.
2) The above cohesive subgraph-based studies often traverse the entire network to build efficient indexes. However, the detected community actually only involves the nodes around the query node, resulting in the traversal of many nodes that are not relevant to the community. 
3) The above studies directly match keywords \cite{Fang_01, Huang_02}, which encounter mismatches due to synonym variations and reduced accuracy from including irrelevant terms. These issues lead to poor semantic cohesiveness in the detected communities.

\textbf{Challenges and our solutions. }To address the above limitations, we face the following three challenges.

\textit{Challenge I: How to simultaneously consider both keywords and locations, as well as the variable travel times between locations, for community detection? }
To suit variable travel times on the road, we use time-dependent spatial cohesiveness to constrain the short travel times between nodes within a community.
Furthermore, we adopt the community Manhattan distance \cite{Wenhua_01} to balance both semantic and time-dependent spatial cohesiveness.

\textit{Challenge II: How to minimize the information accessed in the network? }
We design local methods to gradually expand outward from the query node to detect a $k$-core with cohesive attributes. 
They access only the locally attributed network near the query node, thereby minimizing the network information accessed. 

\textit{Challenge III:  How to address poor accuracy caused by differently expressed synonyms and irrelevant words? }
We utilize Large Language Models (LLMs) \cite{Wei_01} for keyword embedding and calculate the cosine similarity between the two embeddings. 
Since expressed synonyms and irrelevant words make LLMs difficult to capture semantic associations, we design prompts to guide GPT-3.5-Turbo to process keywords before leveraging LLMs for keyword embedding.

\textbf{Contributions.} The contributions are summarized as follows:
\begin{enumerate}[topsep=0pt, partopsep=0pt, itemsep=0pt, parsep=0pt, label=(\arabic*), leftmargin=*]
    \item We propose the problem of identifying semantic-spatial aware $k$-core in time-dependent road-social attributed networks, where the $k$-core containing the query node satisfies high semantic and time-dependent spatial cohesiveness.

    \item We propose two seed expansion algorithms, ESSAC and GSSAC, to mine semantic-spatial aware $k$-core. This is the first work to use local methods to find the $k$-core with cohesive attributes, accessing only the locally attributed network near the query node. 
    
    \item To measure semantic cohesiveness, we design a method based on LLMs like GPT-3.5-Turbo and text-embedding-3-small to calculate semantic similarity between keywords. 

    \item Experimental results show that GSSAC outperforms baselines in terms of structural, semantic, and spatial cohesiveness. 

\end{enumerate}

The rest of the paper is organized as follows. Section \ref{Preliminaries} introduces the $k$-core, A* algorithm, and non-dominated solutions. Section \ref{Problem Formulation} formulates the problem and objective functions. Section \ref{Methods} describes the ESSAC and GSSAC algorithms, along with time-dependent distance and complexity analysis. Section \ref{Experiments} presents the experimental results. Section \ref{Related Work} reviews related studies. Section \ref{Conclusion} concludes the paper.

\section{Preliminaries}\label{Preliminaries}

In this section, we introduce the definition of the $k$-core, the fundamental idea of the A* algorithm, and the concept of non-dominated solutions.

\subsection{{\itshape{k}}-core}\label{k-core}
The $k$-core \cite{Sozio_01} is a widely used dense subgraph model for community search. Given an integer $k$ and an undirected graph $G$, a $k$-core is a connected subgraph $H \subseteq G$, where the degree of each node in $H$ is at least $k$.

\subsection{A* algorithm}\label{A* Algorithm}
Hart et al. \cite{Hart_01} proposed a heuristic strategy based on the A* algorithm to find the shortest path. 
It adopts an actual cost  $g(n)$ from $n_{s}$ to node $n$ and an estimated cost  $h(n)$ from $n$ to $n_{e}$  to compute the total cost $f(n)$ from $n_{s}$ to $n_{e}$.
Based on $f(n)$, the process is as follows:
1) Initially, the strategy maintains an “open” list for nodes that have not been expanded yet and a “close” list for those that have already been expanded. At the start, $n_{s}$ is placed in the “open” list. 
2) The strategy removes the node $n$ with the minimum $f(n)$ value from the “open” list and adds it to the “close” list. 
Then, all neighbors of $n$ are handled as follows. If a neighbor node $m$ is already in the “close” list, it is skipped. If $m$ is not in the open list, it is added to the “open” list. If $m$ is already in the “open” list, the value of $g(m)$ is updated. 
Repeat the second step until $n_{e}$ is added to the “close” list, indicating that the shortest path is found. 

\subsection{Non-dominated solutions}\label{Non-dominated Solutions} 
A non-dominated solution is one that is not dominated by any other solution in the solution space. The dominance relation \cite{Li_02} is adopted to filter out non-dominated solutions in the solution space, defined as follows:

Definition 1 (Dominance relation \cite{Li_02}). Given maximization objective functions $F_{max} = (f_{1}, f_{2}, \dots, f_{n})$ and two solutions $x_{1}, x_{2} \in X$, if $\ \forall i, f_{i}(x_{1}) \le f_{i}(x_{2})$ and $\exists j, f_{j}(x_{1}) < f_{j}(x_{2})$, then $x_{1}$ is dominated by $x_{2}$, denoted: $x_{2} \succ x_{1}$.

Among non-dominated solutions, the preferred solution is found using the Manhattan distance \cite{Wenhua_01}. For two maximization objective functions $f_{1}$ and $f_{2}$, the Manhattan distance $MD_{x}$ for each non-dominated solution $x$ is computed as:

\begin{equation}\label{equ:MD} 
    MD_{x} = f_{1}^{'}(x) + f_{2}^{'}(x)
\end{equation} 
\begin{equation}\label{equ:norm MD} 
    f_{i}^{'}(x) = \frac{f_{i}(x) - f_{i}^{max}}{f_{i}^{min} - f_{i}^{max}} 
\end{equation} 
where $f_{i}(x)$ is the value of solution $x$ in objective function $f_{i}$, and $f_{i}^{'}(x)$ is its normalized value. $f_{i}^{min}(f_{i}^{max})$ denotes the minimum (maximum) value of $f_{i}$. The solution with the minimum MD is the preferred non-dominated solution.

\begin{figure*}[!t]
  \centering
  \includegraphics[width=\linewidth]{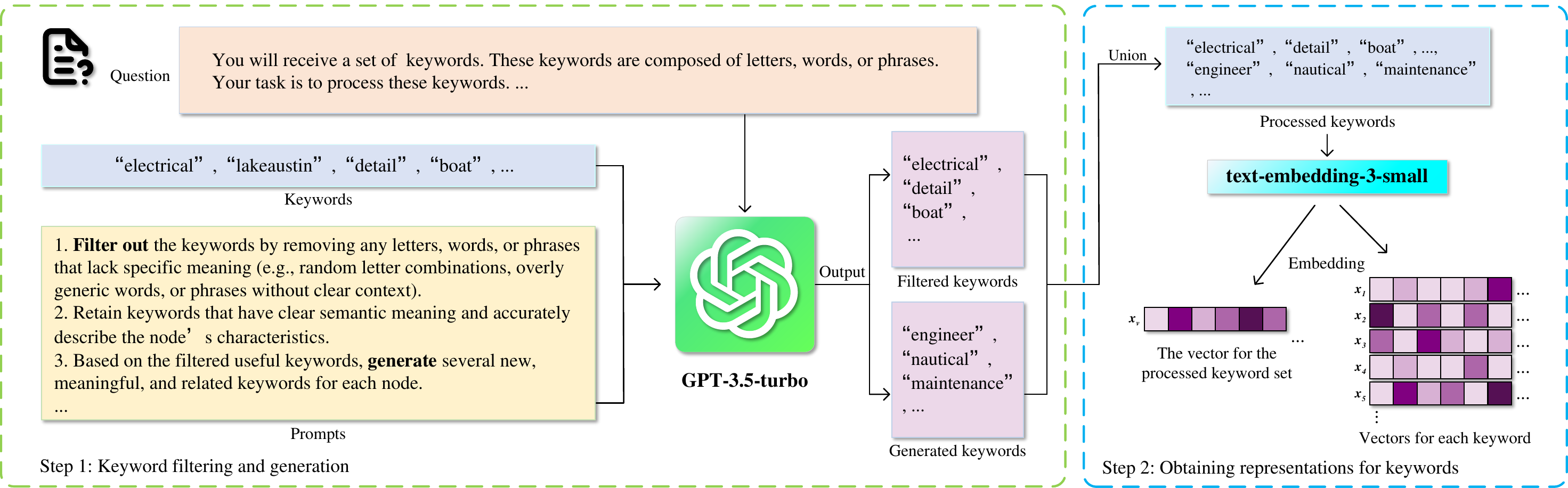}
  \caption{Illustration of 
  the process for generating semantic vectors using LLMs. Prompts guide GPT-3.5-Turbo to process keywords, and text-embedding-3-small is used to obtain their semantic vectors.
  }
  \label{figure:GPT_input_output}
\end{figure*}

\section{Problem Formulation}\label{Problem Formulation}

In this section, we formally introduce the problem and objective functions, followed by a description of the semantic representations.

\subsection{Problem statement}
\noindent  

\indent \textbf{Data model. } The time-dependent road-social attributed network is denoted by $G = (G_{s}, G_{r})$, where $G_{s}$ and $G_{r}$ represent the attributed social network and the time-dependent road network, respectively. The time-dependent road network $G_{r} = (V_{r}, E_{r}, W)$ is a weighted undirected graph, where the node set $V_{r}$ represents the road locations, the edge set $E_{r} = \{(u_{r}, v_{r})|u_{r}, v_{r} \in V_{r} \wedge u_{r}  \ne v_{r} \}$ represents roads, and the weight $W$ represents the travel time of the roads. 
Here, the weight $w_{u_{r}, v_{r}}(t) \in W$ represents the travel time on the road segment $(u_{r}, v_{r}) \in E_{r}$ at time $t$, and varies as a piecewise linear function over time $t$.
The attributed social network $G_{s} = (V_{s}, E_{s}, A, L)$ is an unweighted undirected graph, where $V_{s}$, $E_{s}$, $A$, and  $L$ represent nodes set, edges set, keyword attributes and locations, respectively. 
For a node $v_{s} \in V_{s}$, $attr(v_{s}) \subseteq A$ represents the keyword set of $v_{s}$, and $l_{v_{s}} \in L$ represents the location of $v_{s}$ in the road networks. 
For better understanding, Figure \ref{figure:toy} shows a toy time-dependent road-social attributed network. 

\textbf{Problem 1 } (SSAC search). Given a time-dependent road-social attributed network $G = (G_{s}, G_{r})$ and query information $\left \langle q, k, s_{q}, l_{q}, t \right \rangle $, the objective is to identify a community $c$ that includes the query node $q$ and satisfies the following properties:
\begin{itemize}[leftmargin=*]
    \item \textbf{Structural cohesiveness}: Community $c$ contains the query node $q$ and satisfies $k$-core structure.
    \item \textbf{Semantic cohesiveness}: The keywords of nodes in community $c$ are semantically well-matched with the query keywords $s_{q}$.
    \item \textbf{Time-dependent spatial cohesiveness}: At time $t$, the travel time from the community nodes to location $l_{q}$ in the road network should be as small as possible.  
\end{itemize}

The challenge of addressing the above problem is that semantic and time-dependent spatial cohesiveness may conflict with one another. For instance, while expanding the community, nodes that improve semantic cohesiveness may diminish time-dependent spatial cohesiveness.

\subsection{Objective functions}
To identify SSAC, we design semantic similarity $K_{c}$ and average time-dependent distance $T_{c}$ to measure the semantic and time-dependent spatial cohesiveness of community $c$, calculated as follows:
\begin{align}\label{equ:objective}
    \left\{
    \begin{aligned}
        \max \; & K_{c} = \frac{\bm{\mathit{x_{c}}} \cdot \bm{\mathit{x_{s_{q}}}}}{\| \bm{\mathit{x_{c}}} \|  \| \bm{\mathit{x_{s_{q}}}} \| } + \frac{1}{|c|} \sum_{v \in c} max_{\bm{\mathit{x_{o}}} \in \bm{\mathit{X_{v}}}, \bm{\mathit{x_{p}}} \in \bm{\mathit{X_{s_{q}}}}} \frac{\bm{\mathit{x_{o}}} \cdot \bm{\mathit{x_{p}}}}{\| \bm{\mathit{x_{o}}} \|  \|  \bm{\mathit{x_{p}}} \| }  \\
        \max \; & T_{c} = - \frac{\sum_{v \in c} d_{l_{v}, l_{q}, t}}{|c|}
    \end{aligned}
    \right.
\end{align}
where $\bm{\mathit{x_{s_{q}}}}$  ($\bm{\mathit{x_{c}}}$) is the semantic vector of keywords of community $c$ (query keywords $s_{q}$), $d_{l_{v}, l_{q}, t}$ means the time-dependent distance (Section \ref{sec:Time-dependent distance}) from the location of node $v$ to the location $l_{q}$ at time $t$ calculated by the method based on A* algorithm (Section \ref{A* Algorithm}), $|c|$ is the number of nodes in the community $c$, 
$\bm{\mathit{X_{s_{q}}}}$ ($\bm{\mathit{X_{v}}}$) is the set of semantic vectors for each keyword in $s_{q}$ (each keyword of node $v \in c$).
Specifically, each vector $\bm{\mathit{x_{p}}} \in \bm{\mathit{X_{s_{q}}}}$ ($\bm{\mathit{x_{o}}} \in \bm{\mathit{X_{v}}}$) represents the semantic vector of each keyword in query keywords $s_{q}$ (associated with node $v$).
The first term of $K_{c}$ measures the similarity between the query keywords and the keywords of community $c$ at the community level, while the second term measures the similarity between the query keywords and the keywords of individual nodes within $c$ at the node level. Additionally, $\bm{\mathit{x_{c}}}$, $\bm{\mathit{x_{s_{q}}}}$, $\bm{\mathit{X_{v}}}$, and $\bm{\mathit{X_{s_{q}}}}$ are generated based on large language models (LLMs), as introduced below.

\textbf{Semantic representations. }
Most existing keyword-based community search approaches rely on direct keyword matching \cite{Fang_01, Huang_02}. They fail in two cases: 1) irrelevant, ambiguous, or overly specific terms are included among the keywords; and 2) semantically relevant keywords are expressed in diverse expressions.

To address the above two disadvantages, before leveraging LLMs for keyword embedding, we design prompts to process the keywords. 
Specifically, for the first issue, we design prompts to guide GPT-3.5-Turbo\footnote{https://platform.openai.com/docs/models/gpt-3-5-turbo} in filtering out irrelevant, ambiguous, and overly specific keywords. 
For the second issue, we design prompts for GPT-3.5-Turbo to comprehend the filtered keywords' semantics to generate additional semantically relevant keywords to augment the semantics of the keywords.  
These filtered and generated keywords are then regarded as processed keywords.
Subsequently,  we utilize text-embedding-3-small\footnote{https://platform.openai.com/docs/guides/embeddings/embedding-models}, also developed by OpenAI\footnote{https://openai.com/about/}, to embed the processed keywords to obtain their semantic representations. 

The process is illustrated in Figure \ref{figure:GPT_input_output}. 
The prompts are demonstrated in the left part of Figure \ref{figure:GPT_input_output}. 
The first two prompts aim to filter out semantically irrelevant or ambiguous keywords while retaining those with clear semantic meaning. The third prompt operates on the retained keywords, generating additional semantically relevant keywords. Based on these three prompts, GPT-3.5-Turbo processes the keywords to obtain both filtered and generated keywords, which are then combined to form the “processed keywords”. Subsequently,  we use text-embedding-3-small to embed the processed keywords, obtaining their semantic representations as vectors. For node $v$, its keyword semantic vector is denoted as $\bm{\mathit{x_{v}}}$, and the semantic vectors for each keyword in $attr(v)$ are represented as $\left \{  \bm{\mathit{x_{1}}}, \bm{\mathit{x_{2}}}, \dots \right \}$. Finally, the semantic similarity between keywords is computed using the cosine similarity of their corresponding semantic vectors.

\section{Methods}\label{Methods}
This section introduces an Exact Semantic-Spatial Aware $k$-Core mining algorithm, called ESSAC, and a Greedy Semantic-Spatial Aware $k$-Core mining algorithm, called GSSAC.

\subsection{ESSAC}

\begin{algorithm} [!t]
 \caption{ESSAC}
 \label{Algo:DFS method} 
\SetKwData{Left}{left}
\SetKwData{This}{this}
\SetKwData{Up}{up} 
\SetKwFunction{Union}{Union}
\SetKwFunction{DFS}{DFS} 
\SetKwInOut{Input}{input}
\SetKwInOut{Output}{output}
	\Input{$G=(G_{s}, G_{r})$, a query node $q$, a keywords set $s_{q}$, a location $l_{q}$, the time $t$, and an integer $k$} 
	\Output{Community $c$}
        $C_{can} \leftarrow \emptyset$ \; \label{TEG: line1}
        Execute \DFS{$q$, $G_{s}$, $\emptyset$, $k$, $\emptyset$} to obtain $C_{can}$\; \label{TEG: line2}
        Compute the $MD$ value for each community in $C_{can}$\; \label{TEG: line3}
        $c \gets \arg\min_{c' \in C_{can}} MD_{c'}$\; \label{TEG: line4}
        \textbf{return} $c$\; \label{TEG: line5}

        \BlankLine\BlankLine
        \SetKwProg{Fn}{Procedure}{:}{}
        \Fn{\DFS{$sn$, $G_{s}$, $c_{temp}$, $k$, $proc$} \label{DFS: line6}}{
            $c_{temp} \leftarrow c_{temp} \cup \{sn\}$ \; \label{DFS: line7}
            \If{$\forall \ v \in c_{temp},\  \delta_{c_{temp}}(v) \geq k$ \label{DFS: line8}}{ 
                $C_{can} \leftarrow C_{can} \cup \{c_{temp}\}$\; \label{DFS: line9}
            }
             $proc \leftarrow proc \cup \{sn\}$\; \label{DFS: proc}
            \ForEach{$n$ in $N(sn))$ \label{DFS: line11}}{
                \If{$n \notin proc $ \& $\delta(n) \geq k$ \label{DFS: line12}}{ 
                    \DFS{$n$, $G_{s}$, $c_{temp}$, $k$}\; \label{DFS: line13}
                }
            }
        } 
       
 \end{algorithm}

The main idea of ESSAC is first to use a depth-first search method to obtain all $k$-core structures containing the query node and then select one community from these $k$-cores using community Manhattan distance. 
Algorithm \ref{Algo:DFS method} provides the process of ESSAC (lines \ref{TEG: line1}–\ref{TEG: line5}) and the procedure \DFS (lines \ref{DFS: line6}-\ref{DFS: line12}). 
ESSAC begins by initializing the global variable $C_{can}$ as an empty set (line \ref{TEG: line1}), where $C_{can}$ stores identified $k$-cores.
It first executes the procedure \DFS to identify all $k$-cores containing
the query node $q$ and stores them in $C_{can}$ (line \ref{TEG: line2}). 
To evaluate the semantic and time-dependent spatial cohesiveness of $k$-cores in $C_{can}$, ESSAC evaluates those $k$-cores using Manhattan distance of community $c$, calculated as:
\begin{equation}\label{equ:community MD}
    MD_{c} = K_{c}^{'} + T^{'}_{c}
\end{equation}
where $K_{c}^{'}$ and $T_{c}^{'}$ are the normalized values of $K_{c}$ and $T_{c}$, obtained by normalizing $K_{c}$ and $T_{c}$ in formula (\ref{equ:objective}) using formula (\ref{equ:norm MD}).
Since a lower $MD$ value indicates greater cohesiveness of the $k$-core, the $k$-core with the minimum $MD$ value in $C_{can}$ is selected as the returned community $c$ (lines \ref{TEG: line4}-\ref{TEG: line5}).

\begin{figure*}[!t]
  \centering
  \includegraphics[width=0.95\linewidth]{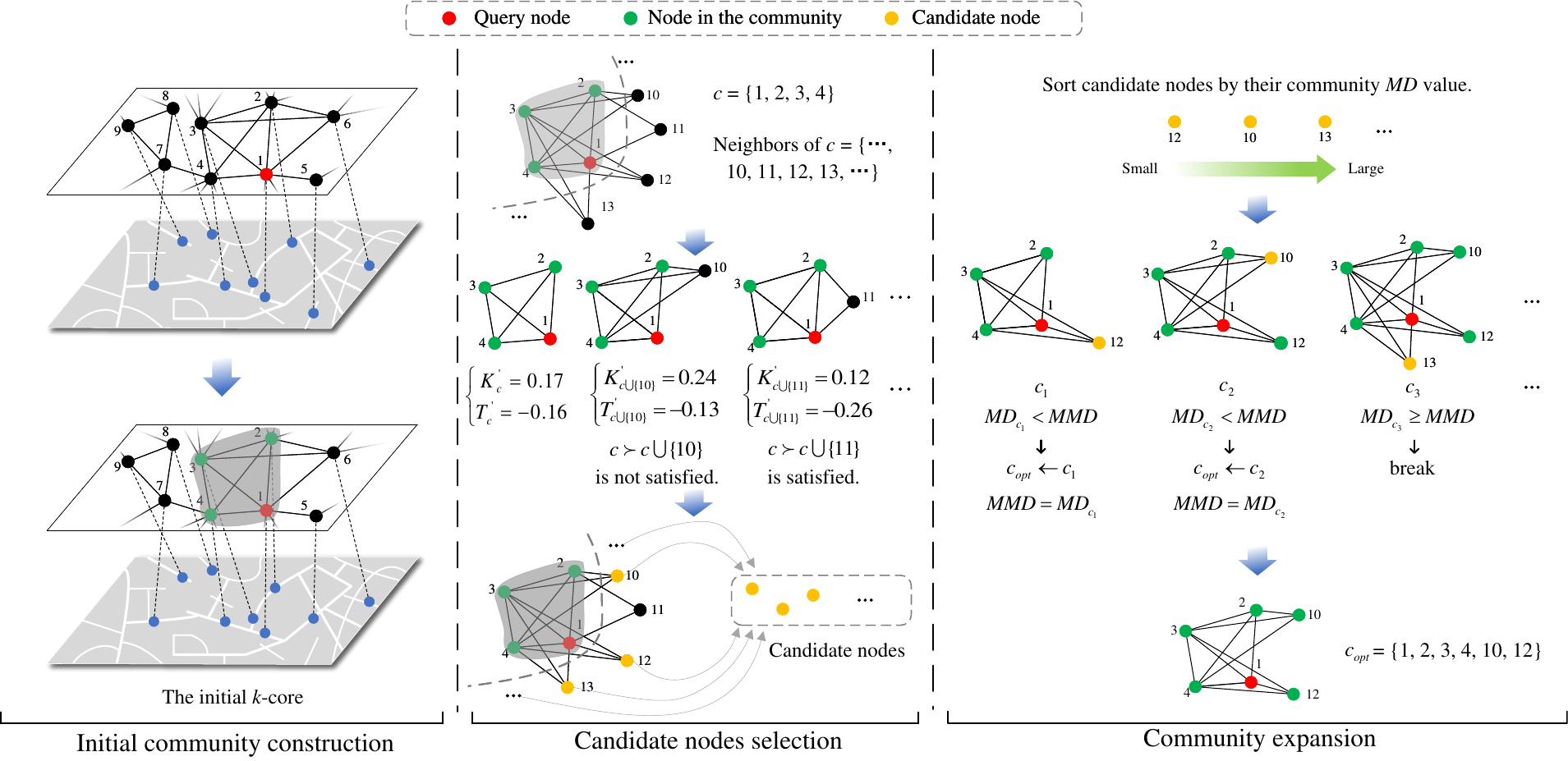}
  \caption{Illustration of GSSAC. It includes initial community construction, candidate nodes selection, and community expansion. } 
  \label{figure:objective_functions}
\end{figure*}

Procedure \DFS aims to identify all $k$-cores containing the query node $q$ and stores them in $C_{can}$.
It takes five parameters: the node to be processed $sn$, the social network $G_{s}$, a current community $c_{temp}$, the integer $k$ of the $k$-core, and a set of processed nodes $proc$. 
\DFS  is a recursive process.
The first time \DFS is called, $sn$ contains the query node $q$, and both $c_{temp}$ and $proc$ are empty (line \ref{DFS: line6}).
Then, the procedure adds $sn$ to $c_{temp}$ and checks whether $c_{temp}$ is a $k$-core (line \ref{DFS: line7}-\ref{DFS: line8}). 
If so, $c_{temp}$ is added to $C_{can}$ (line \ref{DFS: line9}). 
After the above steps, $sn$ is considered a processed node and is added to $proc$ (line \ref{DFS: proc}). 
Next, the neighbor nodes of node $sn$ are processed by calling \DFS in turn (lines \ref{DFS: line11}-\ref{DFS: line13}).
Some neighbor nodes that have already been processed do not need to be processed again, and neighbors with a degree less than $k$ cannot form a $k$-core with $c_{temp}$.
Thus, only unprocessed neighbors with a degree of at least $k$ in $G_{s}$ are processed by \DFS, thereby reducing the number of neighbors that need to be considered (line \ref{DFS: line12}).
After the procedure completes, $C_{can}$ obtains all $k$-cores containing the query node $q$. 
\subsection{GSSAC}
ESSAC  traverses all connected $k$-cores containing the query node. However, many of these $k$-cores exhibit low semantic and time-dependent spatial cohesiveness, i.e., low $K_{c}$ and $T_{c}$ values. 
The possibility of expansion from these low-quality $k$-cores to form highly cohesive communities is low.
To avoid expanding these low-quality $k$-cores, we propose GSSAC, which consistently maintains a single $k$-core with high $K_{c}$ and $T_{c}$ values throughout the expansion process.

GSSAC includes three steps: initial community construction, candidate nodes selection, and community expansion, illustrated in Figure \ref{figure:objective_functions}. 
The first step aims to construct an initial $k$-core with high $K_{c}$ and $T_{c}$ values. The second step identifies candidate nodes that enhance the community's semantic or time-dependent spatial cohesiveness while preserving its $k$-core structure. The third step determines which of these candidate nodes should be added to the community.
For the first step, GSSAC iteratively selects nodes with a high possibility of forming a $k$-core, while maintaining high $K_{c}$ and $T_{c}$ values for the community.
The possibility of node $n$, i.e., $NE_{n}$, calculated as:
\begin{equation}\label{equ:NE}
   NE_{n} = \tilde{y_{1}}(n) + \tilde{y_{2}}(n) - \tilde{d}_{l_{n}, l_{q}, t}(n) 
\end{equation}
where $\tilde{y_{1}}(n)$, $\tilde{y_{2}}(n)$, and $\tilde{d}_{l_{n}, l_{q}, t}(n)$ are the normalized values of $y_{1}(n)$, $y_{2}(n)$, and $d_{l_{n}, l_{q}, t}$ (Section \ref{sec:Time-dependent distance}) using formula (\ref{equ:norm MD}), respectively.
$y_{1}(n)$  and $y_{2}(n)$ are calculated as (\ref{equ:link}) and (\ref{equ:keyword}), respectively.
\begin{equation}\label{equ:link}
    y_{1}(n) = (1 + \frac{1}{(1 + h)^{2}}) |N(n) \cap c|
\end{equation}
where $h$ means the number of hops from node $n$ to query node $q$, $N(n)$ is the neighbors of node $n$ in social network, $|\cdot|$ is the cardinality of a set.
$y_{1}(n)$ indicates the possibility that the community $c$ forms a $k$-core when node $n$ added to $c$.
It increases with more neighbors of $n$ in the community $c$ and fewer hops from node $q$.
\begin{equation}\label{equ:keyword}
    y_{2}(n) = \frac{\bm{\mathit{x_{n}} }\cdot\bm{\mathit{x_{s_{q}}} } }{ {\|\bm{\mathit{x_{n}} }\|} {\|\bm{\mathit{x_{s_{q}}} }\|} }
\end{equation}
where $\bm{\mathit{x_{n}} }$ ($\bm{\mathit{x_{s_{q}}} }$) is the semantic vector of keywords of node $n$ (query keywords $s_{q}$). 
$y_{2}(n)$ represents the semantic similarity between keywords of node $n$ and the query keywords $s_{q}$. 
High $y_{2}(n)$ and -$\tilde{d_{l_{n}, l_{q}, t}}(n)$  suggest that the addition of node $n$ enhances the community's semantic and time-dependent spatial cohesiveness.
A lower $NE_{n}$ value indicates a higher possibility of community $c$ forming a semantic-spatial aware $k$-core when node $n$ is added to $c$. Thus, nodes with the lowest $NE_{n}$ values are progressively added to the community until a $k$-core structure is achieved.

After obtaining the $k$-core structure, the candidate nodes selection step uses community dominance relation to identify candidate nodes that improve the community's cohesiveness. The community dominance relation is defined as follows:

Definition 2 (Community dominance relation \cite{Ni_02}). Given two communities $\{c_{a}, c_{b}\}$ and two maximization objective functions $(f_{1}=K_{c}, f_{2}=T_{c})$, if $\ \forall  i \in \{1, 2\}$, $f_{i}(c_{a}) \le f_{i}(c_{b})$ and if $\ \exists  j \in \{1, 2\}$ such that $f_{j}(c_{a}) < f_{j}(c_{b})$, then $c_{a}$ is dominated by $c_{b}$, denoted as $c_{b} \succ c_{a}$.  

The candidate node improves the community's cohesiveness, meaning that the community formed by adding the candidate node outperforms the original community in at least one of $K_{c}$ or $T_{c}$. In other words, the community formed by adding the candidate node is not dominated by the original community. 
Additionally, the candidate nodes need to satisfy the $k$-core restriction, ensuring that the community remains a $k$-core after their addition. 
In summary, the candidate node $n$ needs to satisfy: (1) $n$ has at least $k$ neighbors within community $c$; (2) $c$ is not dominated by $c\cup\{n\}$.
Among candidate nodes, some nodes enhance the community's semantic cohesiveness, while others improve its time-dependent spatial cohesiveness. Therefore, in the community expansion step, we employ the community $MD$ (refer to formula (\ref{equ:community MD})) to comprehensively evaluate candidate nodes and add those that improve the $K_{c}$ and $T_{c}$ values to the community. GSSAC iteratively executes the second and third steps to expand the community.

\IncMargin{1em}
\begin{algorithm} [!t]
 \caption{GSSAC}
 \label{Algo:The Greedy Method} 
\SetKwData{Left}{left}
\SetKwData{This}{this}
\SetKwData{Up}{up} 
\SetKwFunction{Union}{Union}
\SetKwFunction{Initial}{Initial} 
\SetKwInOut{Input}{input}
\SetKwInOut{Output}{output}

    \Input{$G=(G_{s}, G_{r})$, a query node $q$, a keywords set $s_{q}$, a location $l_{q}$, the time $t$, and an integer $k$} 
    \Output{Community $c$}

    $c \leftarrow \{q\}$; $C_{his}$ $\leftarrow \emptyset$\; \label{step1: line1}
    \While{true \label{step1: line2}}{
         $v \gets \arg\min_{n \in N(c)} \{NE_n \mid \delta(n) \geq k\}$\;\label{step1: line3}
        $c \leftarrow c \cup \{v\}$\;\label{step1: line4}
        \If{ a $k$-core containing $q$ exists \label{step1: line5}}{
            $c \leftarrow$ extract the $k$-core containing $q$\ from $c$\;\label{step1: line6}
            \textbf{break}\;\label{step1: line7}
        }
    }
    \While{true \label{step2: line8}}{
        $N_{can}$ $\leftarrow \emptyset$\; \label{step2: line9}
        \ForEach{$n$ in $N(c)$\label{step2: line10}}{
             \textbf{if} $|N(n)\cap c| \geq k$ \& $c \succ c \cup \{n\}$ is not satisfied\\ \label{step2: line11}
             \quad \textbf{then} $N_{can}$ $\leftarrow$ $N_{can} \cup \{n\}$\;\label{step2: line12}
        }
        \textbf{if} $N_{can}$ {\itshape{is empty}} \textbf{then break}\; \label{step2: line13}
        Compute $MD_{c \cup \{n\}}$ for each $n$ in $N_{can}$\;  \label{step3: line14}
        $N_{sort} \leftarrow $ sort $N_{can}$ by $MD_{c \cup \{n\}}$\; \label{step3: line15}
        $c_{opt} \leftarrow c$; $MMD \leftarrow \infty$\; \label{step3: line16}
        \ForEach{$n$ in $N_{sort}$ \label{step3: line17}}{
            \If{$MD_{c_{opt} \cup \{n\}}$ < MMD \label{step3: line18}}{
                $MMD \leftarrow MD_{c_{opt} \cup \{n\}}$; $c_{opt} \leftarrow c_{opt} \cup \{n\}$\;\label{step3: line19}
            }
            \textbf{else break}\;\label{step3: line20}
        }
        $C_{his} \leftarrow C_{his} \cup \{c_{opt}\}$\; \label{step3: line21}
        Compute $MD$ for each community in $C_{his}$\; \label{step3: line22}
        $c^{*} \gets \arg\min_{c' \in C_{his}} MD_{c'}$\; \label{step3: line23}
        \textbf{if} $c^{*}$ {\itshape{is}} $c_{opt}$ \textbf{then} $c \leftarrow c_{opt}$\; \label{step3: line24}
        \Else{\label{step3: line25}
            $c \leftarrow c^{*}$; \textbf{break}\; \label{step3: line26}
        }  
    }
    \textbf{return} $c$\; \label{step3: line27}

 \end{algorithm}
 \DecMargin{1em}

Algorithm \ref{Algo:The Greedy Method} outlines the steps of GSSAC.
Lines \ref{step1: line1}-\ref{step1: line7} detail the initial community construction step. At the start, community $c$ only contains the query node $q$. The historical community set $C_{his}$ stores the communities after each round of community expansion, initialized as an empty set (line \ref{step1: line1}). $N(c)$ represents neighbor nodes set of community $c$. GSSAC selects the node $v \in N(c)$, whose degree is at least $k$ in the social network, with the smallest $NE$ value (refer to formula (\ref{equ:NE})) (line \ref{step1: line3}). Then, $v$ is added to the community $c$ (line \ref{step1: line4}). Next, applying core decomposition \cite{Batagelj_01}, GSSAC checks if a $k$-core containing the query node $q$ exists (line \ref{step1: line5}). If it does, extract the $k$-core and terminate the loop, indicating the completion of the initial community construction (lines \ref{step1: line6}-\ref{step1: line7}). Otherwise, GSSAC continues to expand community $c$ until it forms a $k$-core containing the query node $q$. Since some nodes do not exist in any $k$-cores, the number of iterations could be set to limit the community expansion to avoid GSSAC traversing the entire network to find these nodes' $k$-cores.

Lines \ref{step2: line8}-\ref{step2: line13} elaborate on the step of candidate node selection. First, the candidate node set $N_{can}$ is initialized as an empty set (line \ref{step2: line9}). For each neighbor node $n$ of $c$, if $n$ has at least $k$ neighbors within $c$, and $c \succ c \cup \{n\}$ is not satisfied, $n$ is added to $N_{can}$ (lines \ref{step2: line10}-\ref{step2: line12}). The nodes in $N_{can}$ are used for community expansion. If $N_{can}$ is empty, the community cannot expand further, so the loop terminates and returns $c$ as the final community (line \ref{step2: line13}); otherwise, the community expansion operation continues.

The step of community expansion is presented in lines \ref{step3: line14}-\ref{step3: line27}. To comprehensively evaluate candidate nodes, for each $n \in N_{can}$, GSSAC first calculates $MD_{c \cup \{n\}}$ of $c \cup \{n\}$ (line \ref{step3: line14}). Then, $N_{can}$ is sorted in ascending order according to $MD_{c \cup \{n\}}$ value to obtain $N_{sort}$ (lines \ref{step3: line15}). $c_{opt}$ records the community formed during this expansion round and is set to $c$. $MMD$ stores the minimum value of $MD$ and is initially assigned a value of infinity (line \ref{step3: line16}). Next, each node $n$ in $N_{sort}$ are sequentially combined with $c_{opt}$ to form a community $c_{opt} \cup \{n\}$, and GSSAC checks whether $MD_{c_{opt} \cup \{n\}}$ is less than $MMD$ (lines \ref{step3: line17}-\ref{step3: line18}). If so, $MMD$ is updated to $MD_{c_{opt} \cup \{n\}}$, and node $n$ is added to $c_{opt}$ (line \ref{step3: line19}); otherwise, stop adding nodes to $c_{opt}$ (line \ref{step3: line20}). Once obtaining $c_{opt}$, put $c_{opt}$ into the historical community set $C_{his}$ and compute $MD$ value for each community in $C_{his}$ (lines \ref{step3: line21}-\ref{step3: line22}). Subsequently, GSSAC obtain community $c^{*}$ with the minimum $MD$ value in $C_{his}$ (line \ref{step3: line23}). 
If $c_{opt}$ still has the possibility of expanding into a more cohesive community, meaning that $c^{*}$ is $c_{opt}$, then $c$ is updated to $c_{opt}$, and GSSAC continues to expand community $c$ (line \ref{step3: line24}). 
Otherwise, $c_{opt}$ is not the most cohesive on $C_{his}$, so no further expansion.
Thus, $c$ is updated to $c^{*}$, and the loop terminates (lines \ref{step3: line25}-\ref{step3: line26}). After the loop is completed, the community $c$ is returned (line \ref{step3: line27}).
In the above process, GSSAC visits only the nodes around the query node, rather than all the nodes in the whole network.

\subsection{Time-dependent distance}\label{sec:Time-dependent distance}
The time-dependent distance $d_{l_{v}, l_{q}, t}$ is the shortest travel time from the start location $l_{v}$ to the end location $l_{q}$ at time $t$.
We use the A* algorithm (Section \ref{A* Algorithm}) to compute $d_{l_{v}, l_{q}, t}$.
The approximate travel time $at_{l_{v}, l_{q}, l_{n}, t}$ from  $l_{v}$ to $l_{q}$ through the location of node $n$ at time $t$ is calculated as follows:
\begin{equation}\label{equ:shortest travel time}
    at_{l_{v}, l_{q}, l_{n}, t} = p(l_{v}, l_{n}, t) + e(l_{n}, l_{q}, t)
\end{equation}
where $p(l_{v}, l_{n}, t)$ denotes the actual shortest travel time from $l_{v}$ to $l_{n}$ at time $t$, and $e(l_{n}, l_{q}, t)$ represents the estimated travel time from $l_{n}$ to $l_{q}$ at time $t$. 
When the A* algorithm processes node $n$ to its neighbor $n+1$,  $p(l_{v}, l_{n+1}, t)$ is computed  as:
\begin{equation}\label{equ:actual shortest travel time}
p(l_{v}, l_{n + 1}, t) = p(l_{v}, l_{n}, t) + w_{l_{n}, l_{n + 1}}(t)
\end{equation}
where $w_{l_{n}, l_{n + 1}}(t)$ represents the travel time between $l_{n}$ and $l_{n + 1}$ at time $t$ on time-dependent road network $G_{r}$.
$e(l_{n}, l_{q}, t)$ is calculated by dividing the Euclidean distance between $l_{n}$ and $l_{q}$ by the road travel velocity at the current time $t$, where the velocity is the ratio of the shortest path distance between $l_{v}$ and $l_{n}$ to the travel time $p(l_{v}, l_{n}, t).$ 
Based on equations (\ref{equ:shortest travel time}) and (\ref{equ:actual shortest travel time}), the A* algorithm is executed. When $l_{n} = l_{q}$ is satisfied, the shortest travel time from $l_{v}$ to $l_{q}$ is found and $d_{l_{v}, l_{q}, t}$ is set to $at_{l_{v}, l_{q}, l_{n}, t}$.

\subsection{Complexity}
The time complexity of GSSAC is $O((n_{l}^{2}d^{2}\log(n_{l}d) + n_{v}n_{s_{q}})\frac{c^{2}}{s} )$, where $n_{l}$ represents the number of average nodes traversed in the road network when calculating time-dependent distance, $d$ is the average degree in the social network, $c$ is the community size, $s$ denotes the average number of nodes added in each expanding iteration, and $n_{v}$ ($n_{s_{q}}$) represents the average number of keywords in node $v \in c$ (in query keywords $s_{q}$).
The reason for the high cost is that the cost of calculating the time-dependent distance is $O((n_{l}^{2}d^{2}\log(n_{l}d))\frac{c^{2}}{s} )$.
The above analysis exclude the cost of utilizing LLMs.

\section{Experiments}\label{Experiments}

In this section, we conduct comprehensive experiments to evaluate the proposed algorithms. We first introduce the experiment setup, including datasets, baselines, implementation details, and evaluation details. Then, we give the experimental results along with relevant analyses.

\subsection{Experimental Setup}
\subsubsection{Datasets. } We use five social networks (Foursquare\_rec\footnote{https://archive.org/details/201309\_foursquare\_dataset\_umn \label{foursquare}}, Foursquare\textsuperscript{\ref{foursquare}}, Weeplace\footnote{http://www.yongliu.org/datasets}, Gowalla\footnote{http://snap.stanford.edu/data/index.html}, and Flickr\footnote{https://www.flickr.com/}) and two road networks (California\footnote{https://users.cs.utah.edu/lifeifei/SpatialDataset.html \label{california}} and North\_America\textsuperscript{\ref{california}}) to construct time-dependent road-social attributed networks, as shown in Table \ref{table:Description of Datasets}. These networks are named by combining the social and road network names, e.g., Foursquare\_rec+CA for the combination of Foursquare\_rec and California. Time-dependent road-social attributed networks are formed by mapping social network nodes, based on their location information, to road network locations. Since Foursquare\_rec, Foursquare, Weeplace, and Gowalla lack keyword information, we supplement them with keywords from DBLP\footnote{http://dblp.uni-trier.de/xml/}. For example, the top 107,092 DBLP node keywords are used to supplement corresponding nodes in Gowalla. For each dataset, we randomly select 200 nodes from the social network, with each node chosen as a query node.
\begin{table}[!t]
  \caption{Statistics of datasets. “$Co_{max}$”, “$Deg_{avg}$”, and “$attr_{avg}$” denote the maximum core number, the average degree of nodes, and the average number of keywords, respectively.}
  \label{table:Description of Datasets}
  \begin{tabular}{p{0.06\columnwidth}<{\centering} p{0.19\columnwidth}<{\centering}p{0.1\columnwidth}<{\centering}p{0.1\columnwidth}<{\centering}p{0.07\columnwidth}<{\centering}p{0.08\columnwidth}<{\centering}p{0.11\columnwidth}<{\centering}}
    \toprule
    Type & Datasets & Nodes & Edges & $Co_{max}$ & $Deg_{avg}$ & $attr_{avg}$ \\
    \midrule
    \multirow{2}{*}{\textit{Road}} & \itshape{California} & 21,048 & 21,693 & 2 & 2.06 & /\\
    \multirow{2}{*}{} & \itshape{North\_America} & 175,813 & 179,102 & 2 & 2.04 & /\\
    \midrule
    \multirow{5}{*}{\textit{Social}} & \itshape{Foursquare\_rec} & 2,551 & 6,469 & 11 & 5.07 & 17.14\\
    \multirow{5}{*}{\textit{}} & \itshape{Foursquare} & 9,100 & 12,090 & 11 & 2.65 & 16.65\\
    \multirow{5}{*}{\textit{}} & \itshape{Weeplace} & 15,795 & 57,046 & 15 & 7.22 & 15.93\\
    \multirow{5}{*}{\textit{}} & \itshape{Gowalla} & 107,092 & 456,830 & 43 & 8.53 & 15.54\\
    \multirow{5}{*}{\textit{}} & \itshape{Flickr} & 214,698 & 2,096,306 & 105 & 19.5 & 13.46\\
  \bottomrule
\end{tabular}
\end{table}

\subsubsection{Baselines. } Baselines include the keyword-based community search method (i.e. ACQ \cite{Fang_01}) and the spatial-aware community search methods (i.e. SLDRG \cite{Ni_01} and LSADEN \cite{Ni_02}).

\subsubsection{Implementation details. } For simplicity, the query keywords and locations for ESSAC and GSSAC are directly taken from the keywords and locations of the query nodes, with $k$ set to 3 and $t$ set to 0. Experiments were conducted on Windows 10 (CPU: Intel(R) Core(TM) i7-4790 @3.60GHz, memory: 16GB) using Python 3.11. 

\subsubsection{Evaluation metrics. } We use {\itshape{coe}} \cite{Duncan_01} (i.e., clustering coefficient)), $GPT_{score}$ \cite{MA_01, jiaan_01}, and $t_{coe}$ to evaluate the community's structural, semantic, and time-dependent spatial cohesiveness, respectively.

Metric $t_{coe}$ is defined as the reciprocal of the maximum travel time from community nodes to the query location, calculated as follows:
\begin{equation}
    t_{coe}(c) = \frac{1}{\max_{v \in c} d_{l_{v},l_{q},t}}
\end{equation}
where $d_{l_{v},l_{q},t}$ represents the shortest travel time from the node $v$ in community $c$ to the query location $l_{q}$ at time $t$. The higher the $t_{coe}$ value indicates that community members are closer to the query location and exhibit greater time-dependent spatial cohesiveness within the community.

\begin{table}[!t]
  \caption{Comparisons of baselines for all given nodes.}
  \label{table:freq_result_LCD}
  \resizebox{\columnwidth}{!}{%
  \begin{tabular}{ccccc}
    \toprule
    Datasets & Metrics & SLDRG & LSADEN & GSSAC \\
    \midrule
    \multirow{3}{*}{\textit{Foursquare\_rec+CA}} & \itshape{coe} & 0.3216 & 0.3409 & \textbf{0.4912} \\
    \multirow{3}{*}{\textit{}} & $GPT_{score}$ & 69.10 & 70.83 & \textbf{79.45}\\
    \multirow{3}{*}{\textit{}} & $t_{coe}$ & 0.0853 & 0.2146 & \textbf{0.2318}\\
    
    \midrule
    \multirow{3}{*}{\textit{Foursquare+CA}} & \itshape{coe} & 0.3346 & 0.3476 & \textbf{0.4785} \\
    \multirow{3}{*}{\textit{}} & $GPT_{score}$ & 70.28 & 71.14 & \textbf{78.47}\\
    \multirow{3}{*}{\textit{}} & $t_{coe}$ & 0.0427 & \textbf{0.0543} & 0.0537\\
   
    \midrule
    \multirow{3}{*}{\textit{Weeplace+CA}} & \itshape{coe} & 0.3372 & 0.3391 & \textbf{0.4638}\\
    \multirow{3}{*}{\textit{}} & $GPT_{score}$ & 70.28 & 71.34 & \textbf{79.43}\\
    \multirow{3}{*}{\textit{}} & $t_{coe}$ & 0.0874 & 0.1327 & \textbf{0.2164}\\
    
    \midrule
    \multirow{3}{*}{\textit{Gowalla+NA}} & \itshape{coe} & 0.3452 & 0.3547 & \textbf{0.4692}\\
    \multirow{3}{*}{\textit{}} & $GPT_{score}$ & 72.32 & 70.69 & \textbf{78.13}\\
    \multirow{3}{*}{\textit{}} & $t_{coe}$ & 0.0372 & 0.0478 & \textbf{0.1364}\\

    \midrule
    \multirow{3}{*}{\textit{Flickr+NA}} & \itshape{coe} & 0.2471 & 0.2867 & \textbf{0.4368}\\
    \multirow{3}{*}{\textit{}} & $GPT_{score}$ & 70.62 & 71.13 & \textbf{79.39}\\
    \multirow{3}{*}{\textit{}} & $t_{coe}$ & 0.0628 & 0.0653 & \textbf{0.0742}\\
    
  \bottomrule
\end{tabular}
}
\end{table}

\subsection{Results}
For SLDRG, LSADEN, and GSSAC, average metrics are calculated for all query nodes. Table \ref{table:freq_result_LCD} shows the average $coe$, $GPT_{score}$, and $t_{coe}$ of GSSAC, SLDRG, and LSADEN for all query nodes. Table \ref{table:freq_result_CS} presents the average $coe$, $GPT_{score}$, and $t_{coe}$ of GSSAC and ACQ for nodes whose communities, as identified by ACQ, are not empty.
Table \ref{table:freq_result_LCD} shows that GSSAC outperforms SLDRG and LSADEN on most datasets. 
In terms of $coe$, GSSAC performs best, followed by LSADEN, and SLDRG performs worst. 
The communities detected by GSSAC are $k$-cores,  ensuring the structural cohesiveness of the communities.
SLDRG and LSADEN are based on local modularity, so their closeness in the community is not as good as the $k$-core.  
In terms of $GPT_{score}$, GSSAC surpasses SLDRG and LSADEN because the latter two methods do not consider keyword information, resulting in weak semantic cohesiveness. As for $t_{coe}$, GSSAC also outperforms SLDRG and LSADEN overall. 
The reason is that GSSAC uses the shortest travel time between two locations in real road networks. 
SLDRG and LSADEN adopt Euclidean distance without considering the real road networks. As a result, communities detected by GSSAC are highly cohesive in the road networks.
Table \ref{table:freq_result_CS} shows that GSSAC outperforms ACQ in structural, semantic, and time-dependent spatial cohesiveness on most datasets. 
Both ACQ and GSSAC achieve comparable structural cohesiveness, as they both use $k$-core as the communities' structure.
GSSAC outperforms ACQ in semantic cohesiveness because ACQ relies on keyword matching, which is affected by redundant or irrelevant words and synonyms in different forms. In contrast, GSSAC uses large language models to process keywords and calculate semantic similarity, alleviating the aforementioned issues.
Additionally, GSSAC surpasses ACQ in time-dependent spatial cohesiveness, as ACQ does not consider location information, resulting in low time-dependent spatial cohesiveness.

\begin{table}[!t]
  \caption{Comparisons of baselines for given nodes whose communities detected by ACQ are not empty. “Num” means the number of nodes whose communities detected by ACQ are not empty.}
  \label{table:freq_result_CS}
  \fontsize{7}{8}\selectfont 
  \resizebox{\columnwidth}{!}{%
  \begin{tabular}{ccccc}
    \toprule
    Datasets & Metrics & ACQ & GSSAC &  Num\\
    \midrule
    \multirow{3}{*}{\textit{Foursquare\_rec+CA}} & \itshape{coe} & 0.4314 & \textbf{0.4767}\\
    \multirow{3}{*}{\textit{}} & $GPT_{score}$ & \textbf{80.54} & 80.12 & 62\\
    \multirow{3}{*}{\textit{}} & $t_{coe}$ & 0.0712 & \textbf{0.2517}\\
    
    \midrule
    \multirow{3}{*}{\textit{Foursquare+CA}} & \itshape{coe} & 0.4256 & \textbf{0.4713}\\
    \multirow{3}{*}{\textit{}} & $GPT_{score}$ & 79.74 & \textbf{81.32} & 96\\
    \multirow{3}{*}{\textit{}} & $t_{coe}$ & 0.0264 & \textbf{0.0523} \\
   
    \midrule
    \multirow{3}{*}{\textit{Weeplace+CA}} & \itshape{coe} & 0.4389 & \textbf{0.4822} \\
    \multirow{3}{*}{\textit{}} & $GPT_{score}$ & 76.14 & \textbf{79.18} & 47\\
    \multirow{3}{*}{\textit{}} & $t_{coe}$ & 0.0434 & \textbf{0.1753} \\
    
    \midrule
    \multirow{3}{*}{\textit{Gowalla+NA}} & \itshape{coe} & 0.4412 & \textbf{0.4734} \\
    \multirow{3}{*}{\textit{}} & $GPT_{score}$ & 77.42 & \textbf{79.48} & 74\\
    \multirow{3}{*}{\textit{}} & $t_{coe}$ & 0.0215 & \textbf{0.1147} \\

    \midrule
    \multirow{3}{*}{\textit{Flickr+NA}} & \itshape{coe} & 0.4218 & \textbf{0.4337} \\
    \multirow{3}{*}{\textit{}} & $GPT_{score}$ & 79.03 & \textbf{80.74} & 73\\
    \multirow{3}{*}{\textit{}} & $t_{coe}$ & 0.0283 & \textbf{0.0659} \\
    
  \bottomrule
\end{tabular}
}
\end{table}

\begin{figure}[!t]
\centering
    \captionsetup[subfigure]{skip=0pt}
    \hspace{10mm}  
    \begin{subfigure}[b]{0.49\columnwidth} 
        \centering
        \captionsetup{belowskip=-6pt, aboveskip=-6pt}  
        \includegraphics[width=\linewidth]{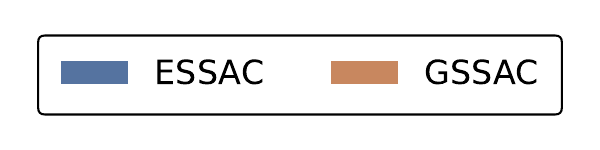}
        \caption{}
        \label{fig:methods}
    \end{subfigure}

    \begin{subfigure}[b]{\columnwidth}
        \includegraphics[width=\linewidth]{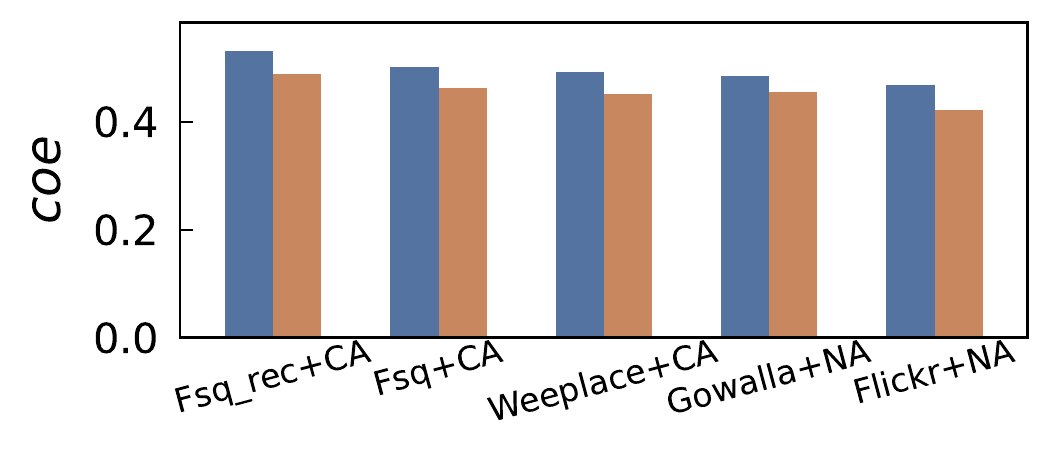}
        \caption{\hspace{0.5cm}(a) $coe$}
        \label{fig:comparison_sub1}
    \end{subfigure}
    \hfill 
    \begin{subfigure}[b]{\columnwidth}
        \includegraphics[width=\linewidth]{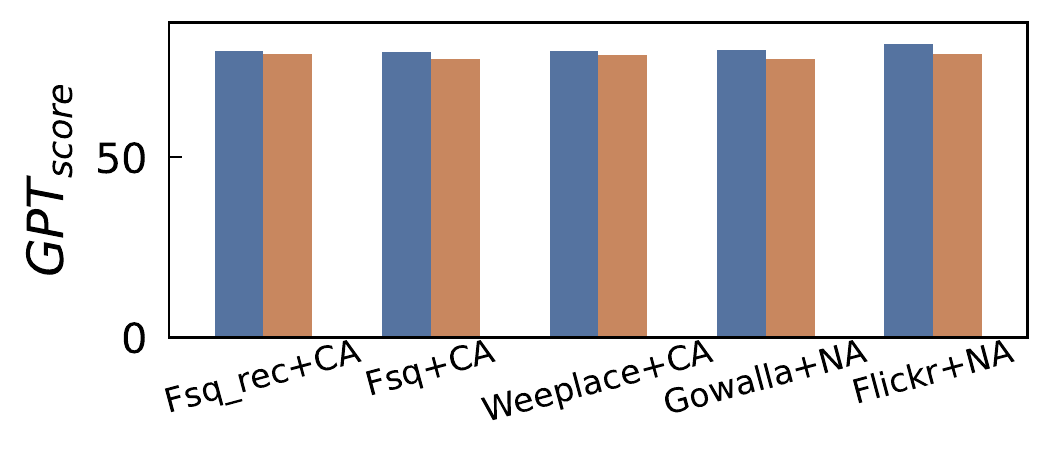}
        \caption{\hspace{0.3cm}(b) $GPT_{score}$}
        \label{fig:comparison_sub2}
    \end{subfigure}
    \hfill  
    \begin{subfigure}[b]{\columnwidth}
        \includegraphics[width=\linewidth]{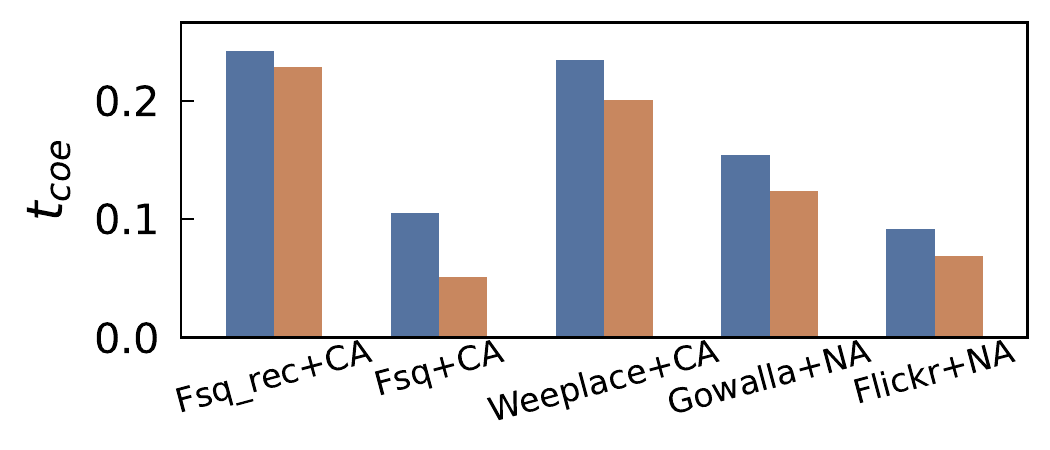}
        \caption{\hspace{0.5cm}(c) $t_{coe}$}
        \label{fig:comparison_sub3}
    \end{subfigure}
    \hfill  
    \begin{subfigure}[b]{\columnwidth}
        \includegraphics[width=\linewidth]{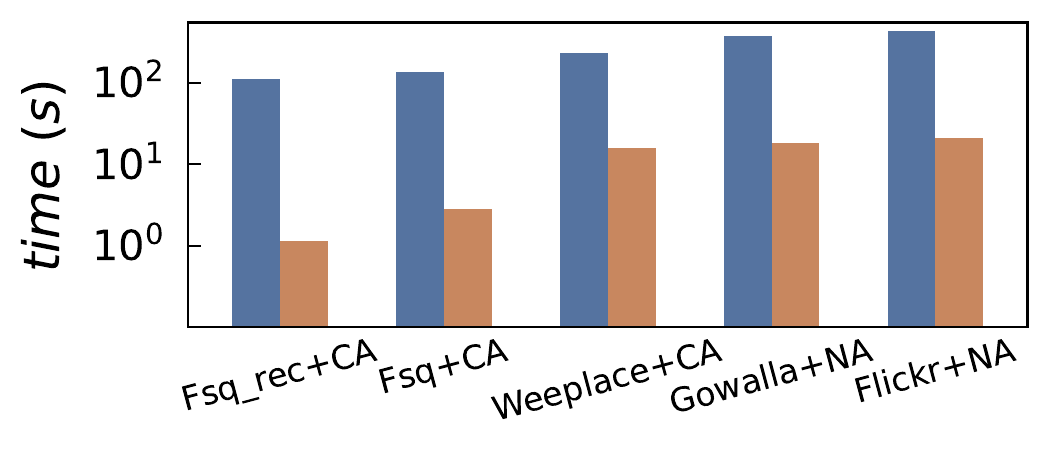}
        \caption{\hspace{0.5cm} (d) time}
        \label{fig:comparison_sub4}
    \end{subfigure}
    
    \caption{Comparison of ESSAC and GSSAC. For simplicity, the abbreviations \textit{Fsq\_rec} and \textit{Fsq} denote the datasets \textit{Foursquare\_rec} and \textit{Foursquare} and are used as such in the following figures and tables. }
    \label{figure:comparison}
\end{figure}

\subsection{Discussion}

\subsubsection{Comparison of ESSAC and GSSAC}
We extract 500-node subgraphs from Foursquare\_rec+CA, Foursquare+CA, Weeplace+CA, Gowalla+NA and Flickr+NA, to test both algorithms. For each dataset, 100 nodes are randomly selected from each subgraph as query nodes. Figure \ref{figure:comparison} shows the results of ESSAC and GSSAC.
It shows that GSSAC performs comparably to ESSAC in structural, semantic, and time-dependent spatial cohesiveness but is nearly 30 times faster in runtime.
ESSAC traverses all $k$-cores containing the query node to find a community, whereas GSSAC maintains a single cohesive community throughout the expansion process.
Most $k$-cores traversed by ESSAC are not cohesive, causing unnecessary computations, whereas GSSAC improves efficiency by focusing on one cohesive community.

\subsubsection{Effect of $k$}
\begin{figure*}[!t]
\centering
    \captionsetup[subfigure]{skip=0pt}
    \hspace{10mm}  
    \begin{subfigure}[b]{0.7\textwidth} 
        \centering
        \captionsetup{belowskip=-4pt, aboveskip=-4pt}  
        \includegraphics[width=\linewidth]{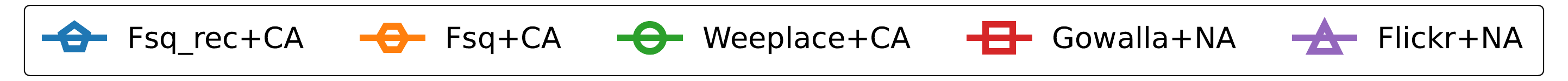}
        \caption{}
        \label{fig:datasets}
    \end{subfigure}

    \begin{subfigure}[b]{0.32\textwidth}
        \includegraphics[width=\linewidth]{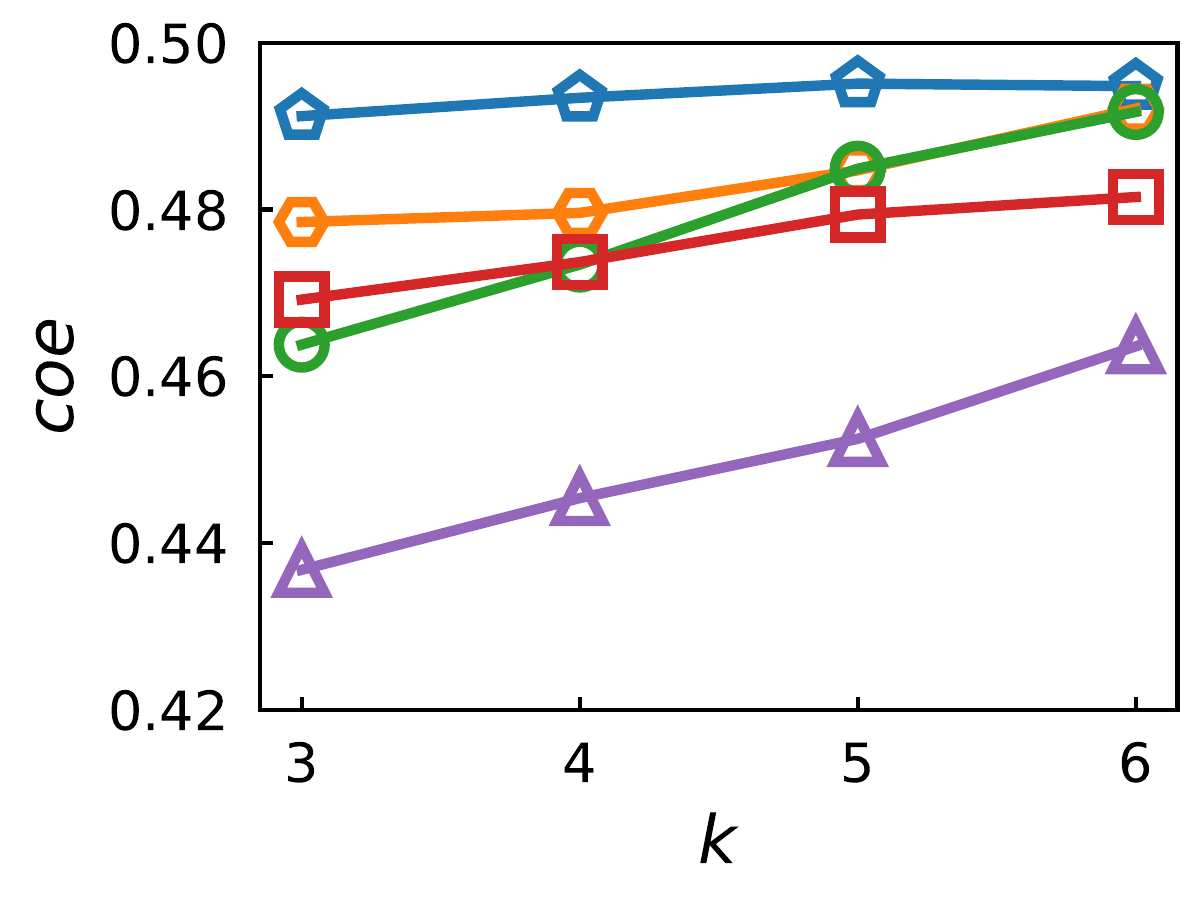}
        \caption{\hspace{0.7cm} (a) $coe$}
        \label{fig:k_sub1}
    \end{subfigure}
    \hfill  
    \begin{subfigure}[b]{0.32\textwidth}
        \includegraphics[width=\linewidth]{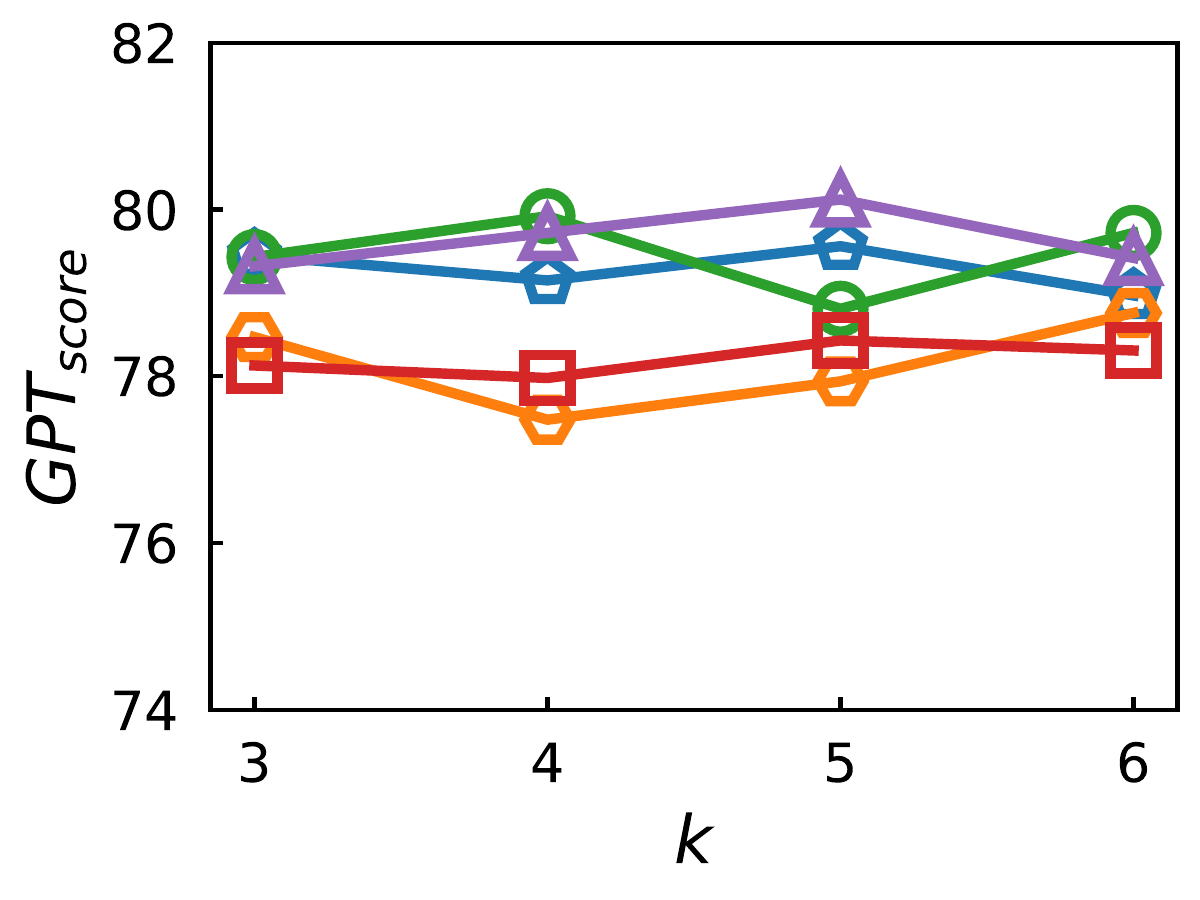}
        \caption{\hspace{0.7cm}(b) $GPT_{score}$}
        \label{fig:k_sub2}
    \end{subfigure}
    \hfill
    \begin{subfigure}[b]{0.32\textwidth}
        \includegraphics[width=\linewidth]{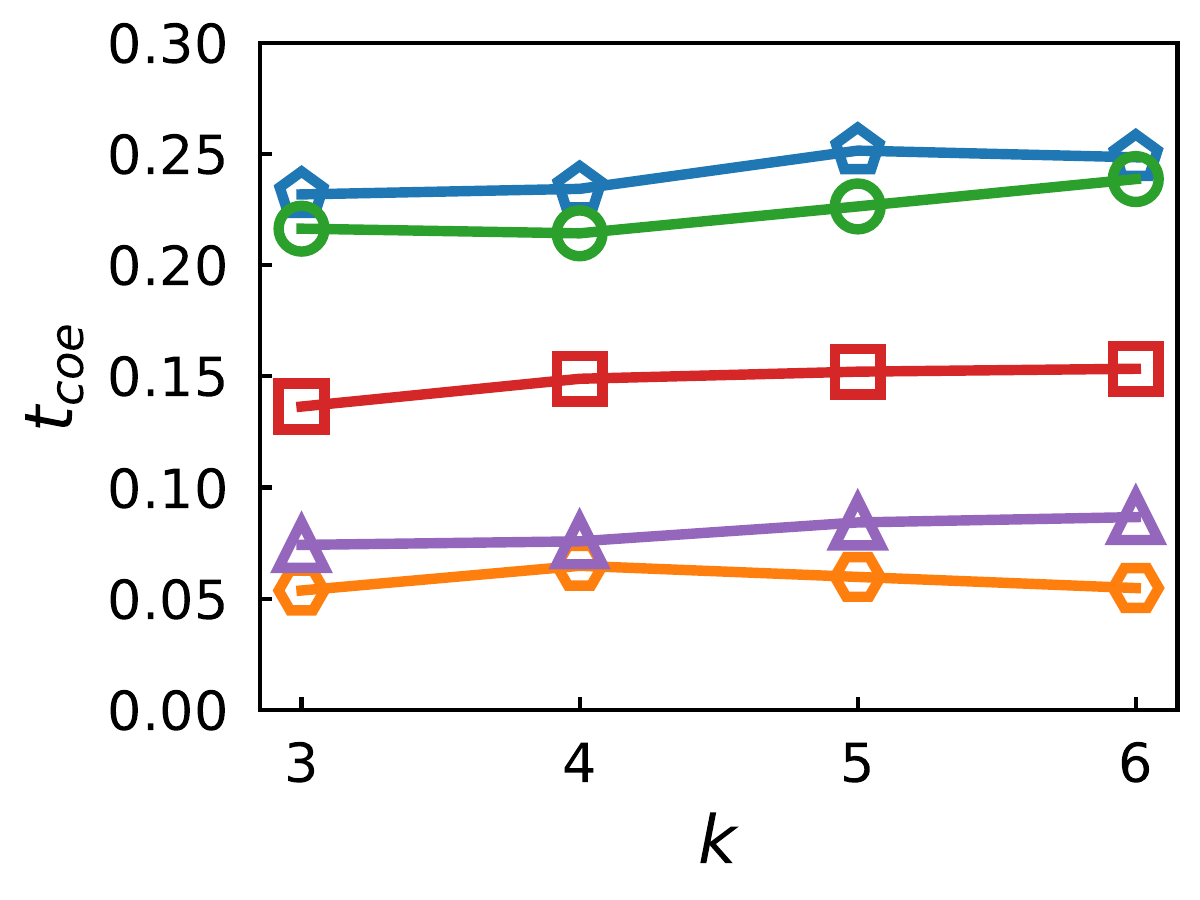}
        \caption{\hspace{0.7cm}(c) $t_{coe}$}
        \label{fig:k_sub3}
    \end{subfigure}
    
    \caption{Results of GSSAC with different $k$. }
    \label{figure:parameters_k}
\end{figure*}

To evaluate the effect of the $k$ parameter on GSSAC's performance, we conducted experiments on the datasets of Foursquare\_rec+CA, Foursquare+CA, Weeplace+CA, Gowalla+NA, and Flickr+NA, with $k$ ranging from 3 to 6. For each dataset, 200 nodes were randomly selected as query nodes.
Figure \ref{figure:parameters_k} shows that as the $k$ value increases, the structural cohesiveness of the community improves since the minimum degree of the nodes increases, leading to stronger connectivity within the community.
The semantic cohesiveness slightly increases and fluctuates as $k$ increases.
time-dependent spatial cohesiveness is also improved as $k$ rises, as the smaller community size brings nodes closer together spatially.

\begin{figure}[!t]
\centering
    \captionsetup[subfigure]{skip=0pt}
    \hspace{7mm}  
    \begin{subfigure}[b]{0.7\columnwidth} 
        \centering
        \captionsetup{belowskip=-3pt, aboveskip=-3pt}  
        \includegraphics[width=\linewidth]{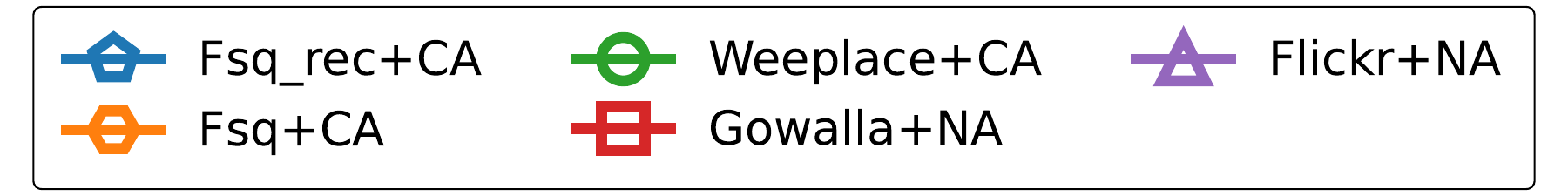}
        \caption{}
        \label{fig:datasets2}
    \end{subfigure}

    \begin{subfigure}[b]{0.48\columnwidth}
        \includegraphics[width=\linewidth]{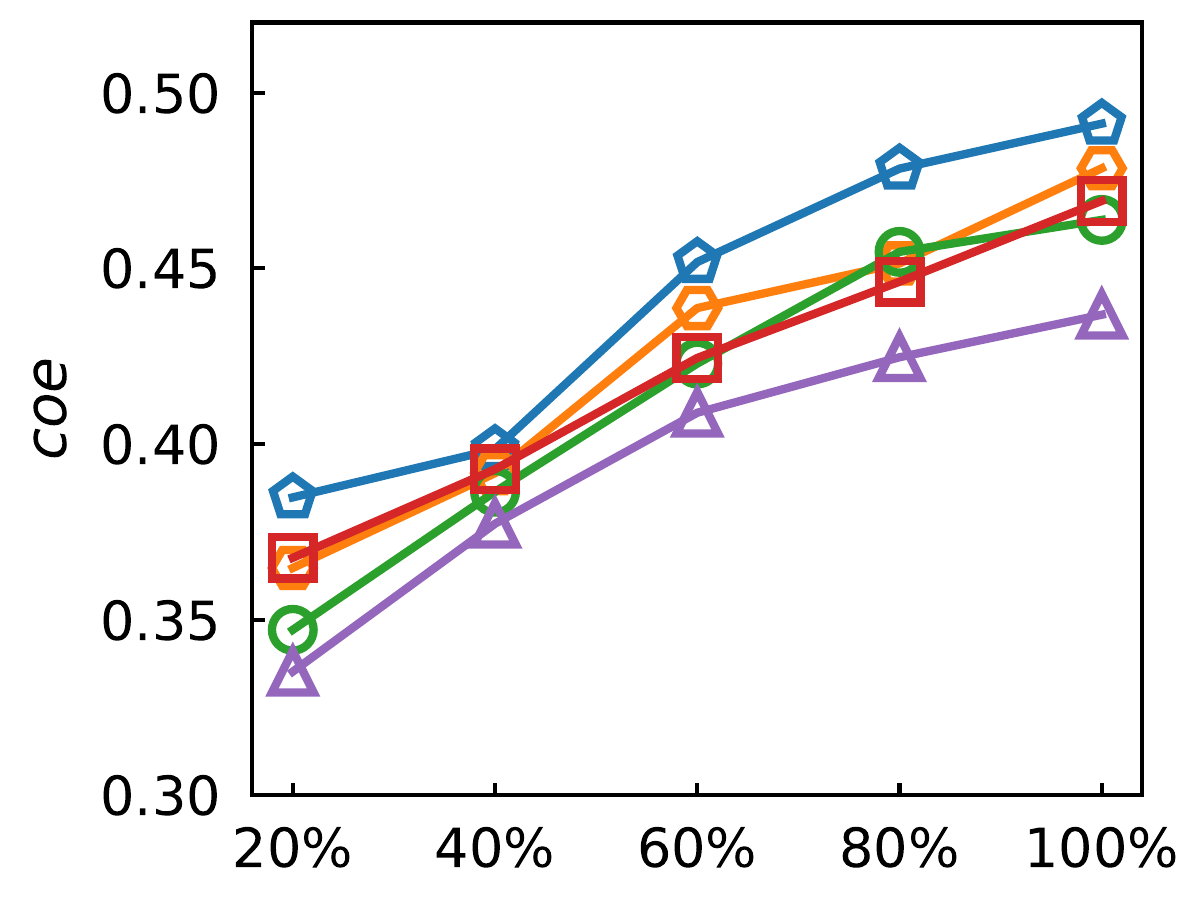}
        \caption{\hspace{0.65cm}(a) $coe$}
        \label{fig:scalability_sub1}
    \end{subfigure}
    \hfill  
    \begin{subfigure}[b]{0.48\columnwidth}
        \includegraphics[width=\linewidth]{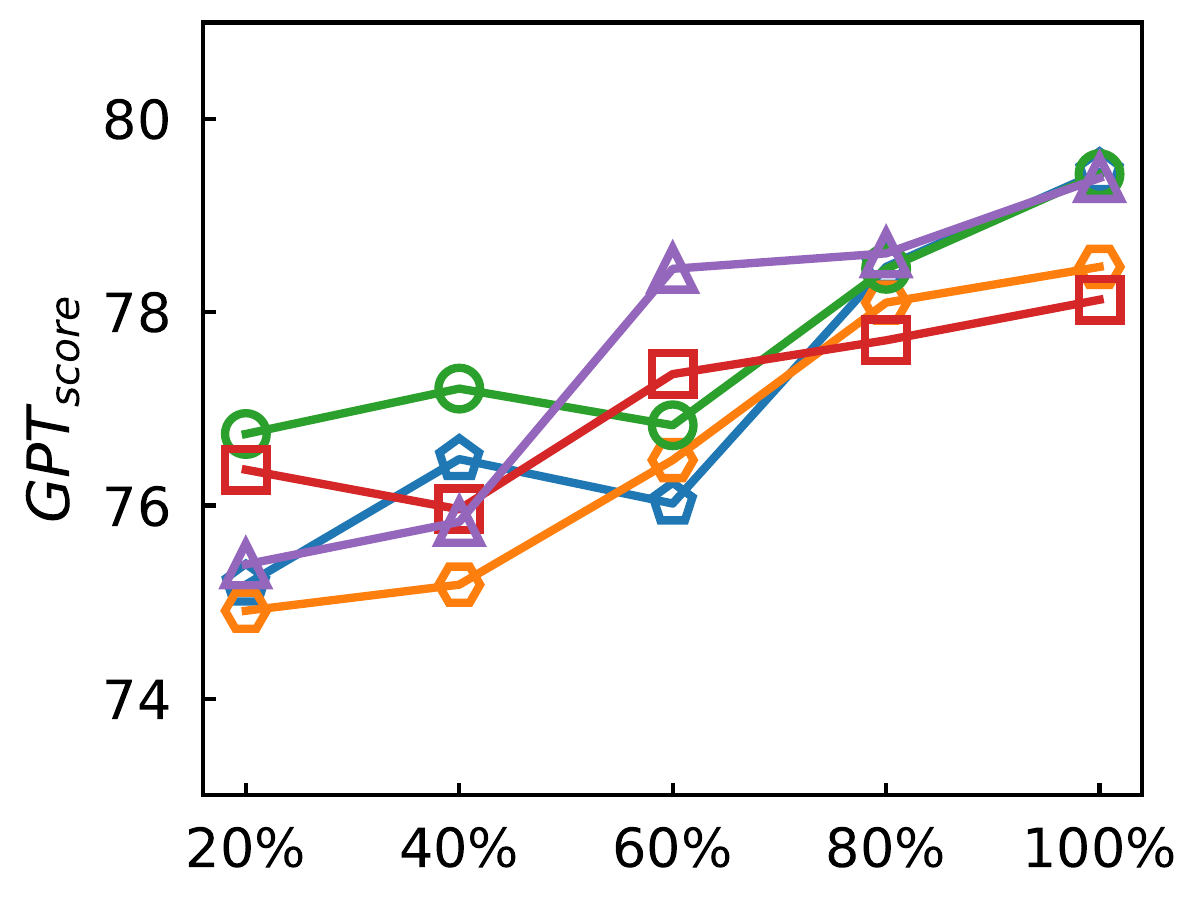}
        \caption{\hspace{0.53cm}(b) $GPT_{score}$}
        \label{fig:scalability_sub2}
    \end{subfigure}
     \medskip 
    \begin{subfigure}[b]{0.48\columnwidth}
        \includegraphics[width=\linewidth]{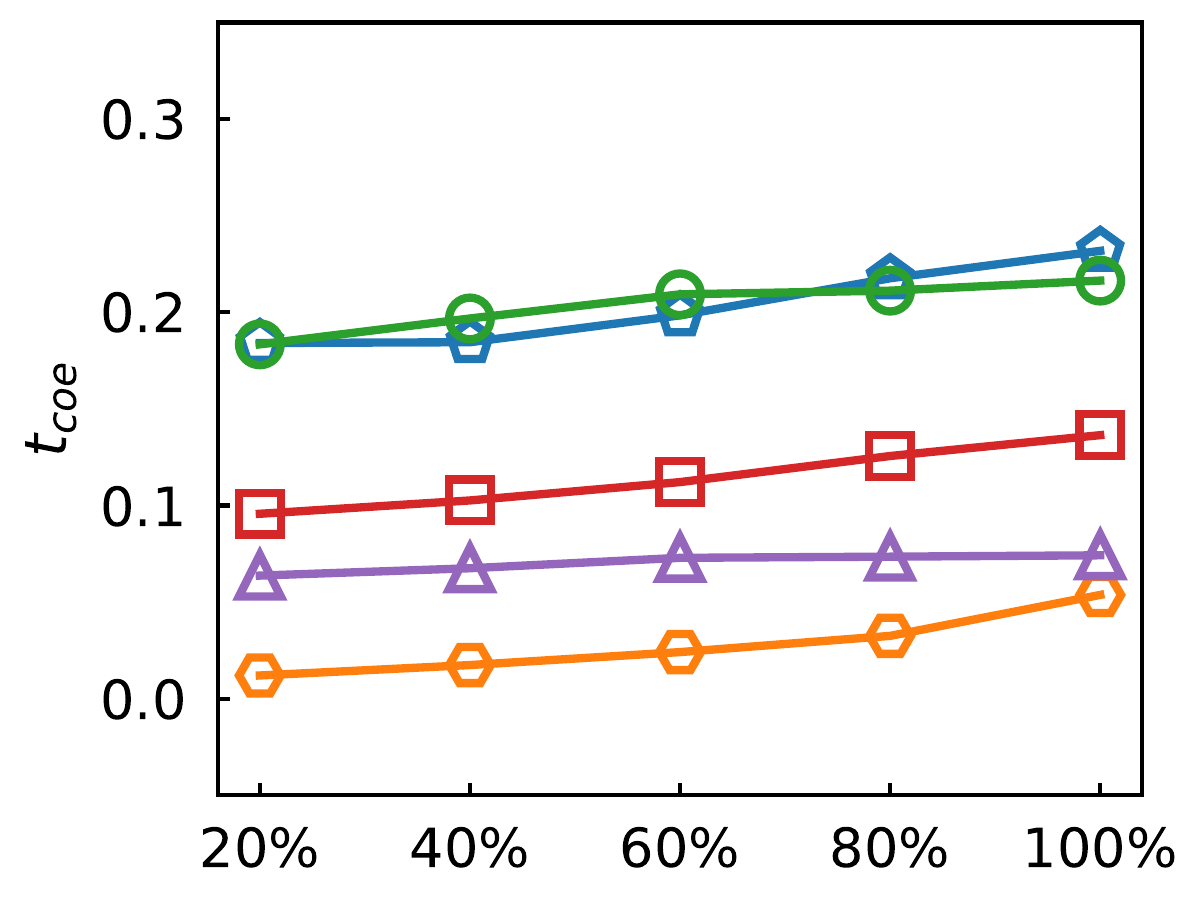}
        \caption{\hspace{0.65cm}(c) $t_{coe}$}
        \label{fig:scalability_sub3}
    \end{subfigure}
    \hfill  
    \begin{subfigure}[b]{0.48\columnwidth}
        \includegraphics[width=\linewidth]{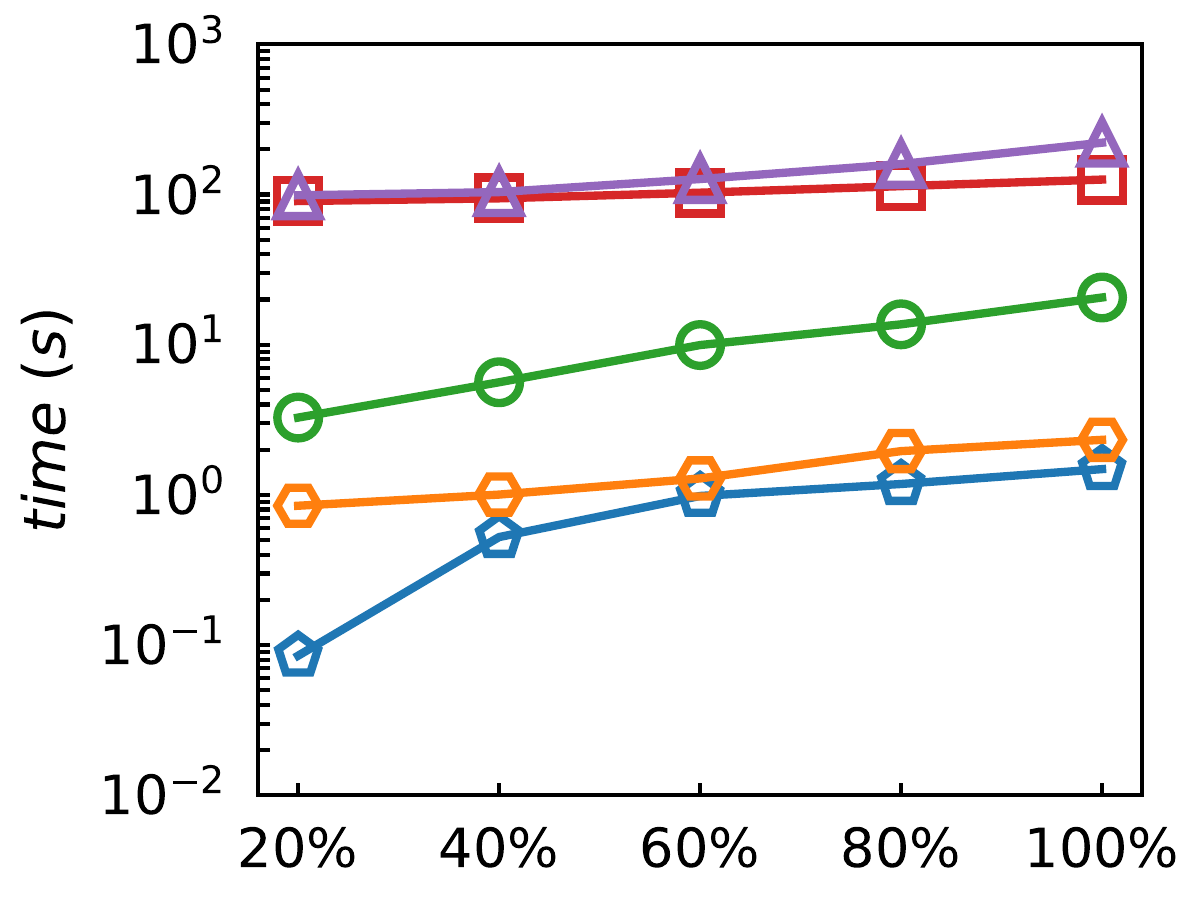}
        \caption{\hspace{0.53cm}(d) time}
        \label{fig:scalability_sub4}
    \end{subfigure}
    \caption{Scalability of GSSAC. }
    \label{figure:scalability}
\end{figure}

\subsection{Scalability}
To test GSSAC's scalability, we randomly extract subgraphs containing 20\%, 40\%, 60\%, 80\%, and 100\% of the nodes from the Foursquare\_rec+CA, Foursquare+CA, Weeplace+CA, Gowalla+NA, and Flickr+NA datasets. Figure \ref{figure:scalability} shows the results for GSSAC on subgraphs of different sizes.
As the subgraph size increases, the structural cohesiveness improves due to denser subgraphs.
Similarly, as more nodes are added to the subgraphs, both the semantic and time-dependent spatial cohesiveness of the communities increase, as the added nodes improve semantic cohesion and become more spatially clustered.
Finally, as the subgraph size increases, the average running time rises slightly, but not significantly, as GSSAC only traverses the local network around the query node, avoiding excessive time consumption.

\begin{table}[!t]
  \caption{Results of ablation study. The metric $G_{score}$ denotes $GPT_{score}$. }
  \label{table:ablation}
  \begin{tabular}{p{0.105\columnwidth}<{\centering} p{0.075\columnwidth}<{\centering}p{0.073\columnwidth}<{\centering}p{0.073\columnwidth}<{\centering}p{0.073\columnwidth}<{\centering}p{0.073\columnwidth}<{\centering}p{0.073\columnwidth}<{\centering}p{0.088\columnwidth}<{\centering}}
    \toprule
    Datasets & Metrics & w/o. icc &w/o. cns & w/o. ce & w/o. GPT & w/o. emb & GSSAC \\
    \midrule
    \multirow{3}{*}{\makecell{\textit{Fsq\_rec} \\ \textit{+}  \\ \textit{CA}}} & \itshape{coe} & 0.3419 & 0.4025 & 0.4481 & 0.4726 & 0.4841  & \textbf{0.4912}\\
    \multirow{3}{*}{\textit{}} & $G_{score}$ & 74.17 & 76.21 & 78.50 & 76.13 & 72.86 & \textbf{79.45} \\
    \multirow{3}{*}{\textit{}} & $t_{coe}$ & 0.1318 & 0.1446 & 0.1925 & 0.2031 & 0.2085 & \textbf{0.2318} \\
    \midrule

    \multirow{3}{*}{\makecell{\textit{Fsq} \\ \textit{+}  \\ \textit{CA}}} & \itshape{coe} & 0.3165 & 0.3817 & 0.4074 & 0.4392 & 0.4438  & \textbf{0.4785}\\
    \multirow{3}{*}{\textit{}} & $G_{score}$ & 73.51 & 74.28 & 76.89 & 75.77 & 73.17 & \textbf{78.47} \\
    \multirow{3}{*}{\textit{}} & $t_{coe}$ & 0.0286 & 0.0301 & 0.0318 & 0.0410 & 0.0421 & \textbf{0.0537} \\
    \midrule

    \multirow{3}{*}{\makecell{\textit{Weeplace} \\ \textit{+}  \\ \textit{CA}}} & \itshape{coe} & 0.3116 & 0.3924 & 0.4247 & 0.4518 & 0.4543  & \textbf{0.4638}\\
    \multirow{3}{*}{\textit{}} & $G_{score}$ & 75.13 & 76.65 & 77.24 & 76.18 & 72.41 & \textbf{79.43} \\
    \multirow{3}{*}{\textit{}} & $t_{coe}$ & 0.1027 & 0.1395 & 0.1721 & 0.1829 & 0.1854 & \textbf{0.2164} \\
    \midrule

    \multirow{3}{*}{\makecell{\textit{Gowalla} \\ \textit{+}  \\ \textit{NA}}} & \itshape{coe} & 0.3017 & 0.3820 & 0.4318 & 0.4485 & 0.4586 & \textbf{0.4692} \\
    \multirow{3}{*}{\textit{}} & $G_{score}$ & 75.26 & 75.46 & 76.53 & 74.43 & 71.21 & \textbf{78.13} \\
    \multirow{3}{*}{\textit{}} & $t_{coe}$ & 0.0459 & 0.0762 & 0.0925 & 0.1109 & 0.1012 & \textbf{0.1364} \\
    \midrule

    \multirow{3}{*}{\makecell{\textit{Flickr} \\ \textit{+}  \\ \textit{NA}}} & \itshape{coe} & 0.2149 & 0.3241 & 0.4146 & 0.4364 & 0.4281 & \textbf{0.4368} \\
    \multirow{3}{*}{\textit{}} & $G_{score}$ & 76.62 & 77.18 & 78.14 & 76.82 & 73.36 & \textbf{79.39} \\
    \multirow{3}{*}{\textit{}} & $t_{coe}$ & 0.0556 & 0.0513 & 0.0651 & 0.0719 & 0.0692 & \textbf{0.0742} \\
    
  \bottomrule
\end{tabular}
\end{table}

\subsection{Ablation study}
We conduct ablation experiments on Foursquare\_rec+CA, Foursquare+CA, Weeplace+CA, Gowalla+NA, and Flickr+NA datasets. Five simplified versions of GSSAC are designed as:
1) “w/o. icc”: Removes the initial community construction step, randomly selecting $k$ nodes from the query node's neighbors.
2) “w/o. cns”: Omits the candidate node selection step, randomly choosing neighbor nodes as candidates.
3) “w/o. ce”: Removes the community expansion step, leaving only the initial community construction.
4) “w/o. GPT”: Uses direct keyword embedding instead of GPT-3.5-Turbo for keyword processing.
5) “w/o. emb”: Replaces embeddings with keyword matching instead of using the text-embedding-3-small model.
For each dataset, 200 query nodes are randomly selected for the ablation experiments. 

The table \ref{table:ablation} presents the results.
It shows that GSSAC outperforms w/o. icc, w/o. cns, w/o. ce, w/o. GPT, and w/o. emb, highlighting the importance of initial community construction, candidate node selection, community expansion, and LLM-based keyword processing in improving community cohesiveness.
The $coe$ of w/o. icc in the Weeplace+CA is 0.3116, much lower than the 0.4638 achieved by GSSAC, indicating the crucial role of the initial community construction step.
Similarly, the $coe$ of w/o. ce in the Weeplace+CA dataset is 0.4247, slightly worse than GSSAC, suggesting that the initial community could achieve high cohesiveness.
w/o. cns perform worse than both w/o. ce and GSSAC, showing that random node expansion leads to poor cohesiveness, emphasizing the importance of candidate node selection.
w/o. emb performs much worse than w/o. GPT in terms of $GPT_{score}$, indicating that keyword embedding using text-embedding-3-small is critical for enhancing semantic cohesiveness.
Finally, the $GPT_{score}$ for w/o. GPT on the Gowalla+NA is 74.43, compared to 78.13 for GSSAC, confirming that the keyword process based on GPT-3.5-Turbo enhances community semantic cohesiveness.

\subsection{Case study}
\begin{figure}[!t]
\centering

    \begin{subfigure}[b]{0.49\columnwidth}
        \includegraphics[width=1\linewidth]{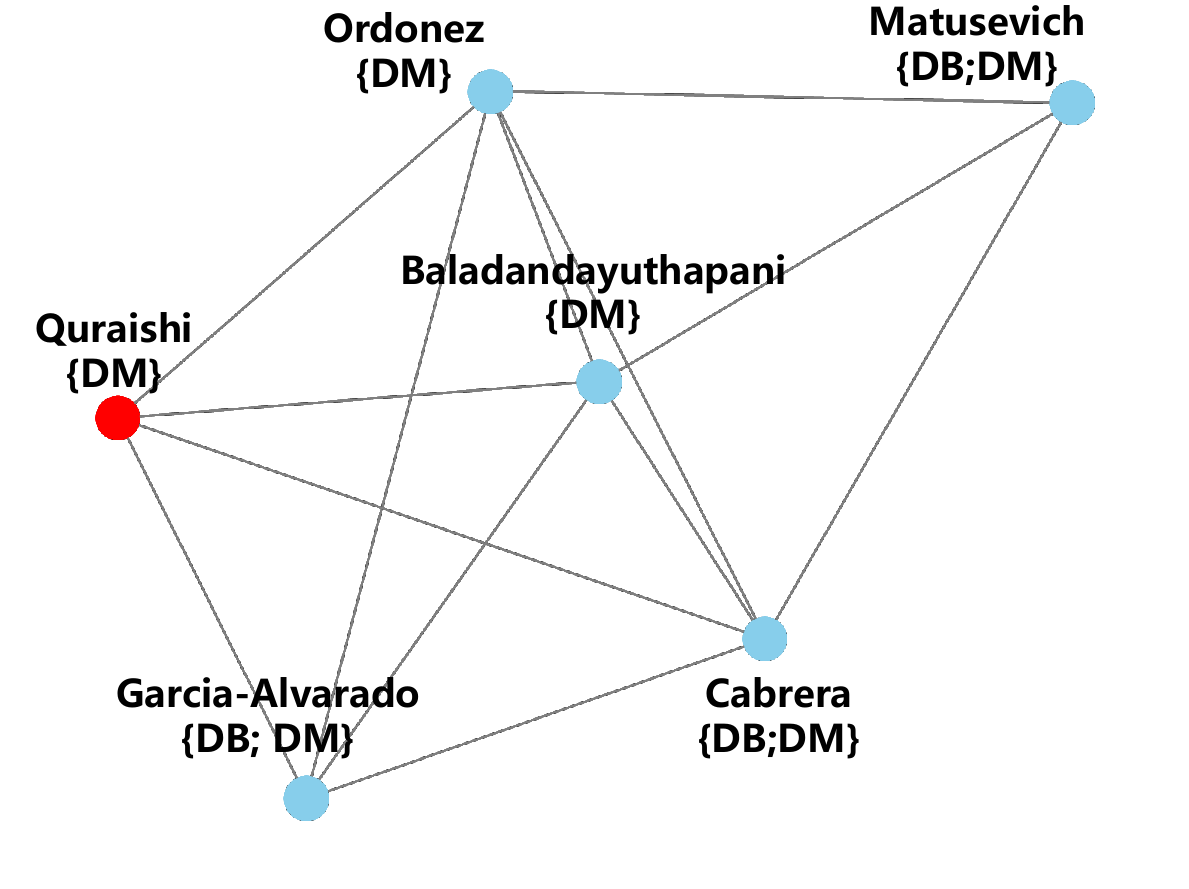}
        \caption{(a) GSSAC}
        \label{fig:case_01}
    \end{subfigure}
    \begin{subfigure}[b]{0.49\columnwidth}
        \includegraphics[width=1\linewidth]{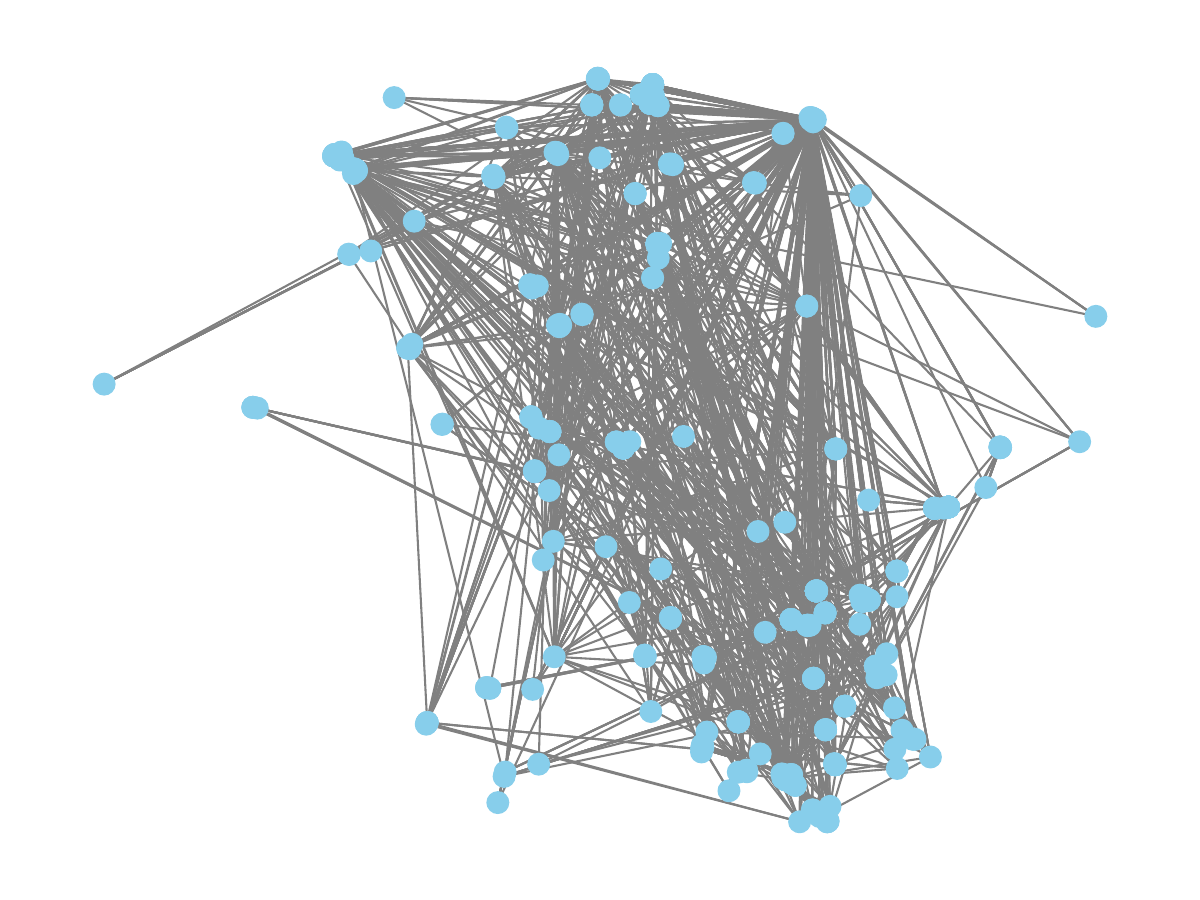}
        \caption{(b) ACQ}
        \label{fig:case_02}
    \end{subfigure}
    \smallskip
    \begin{subfigure}[b]{0.49\columnwidth}
        \includegraphics[width=1\linewidth]{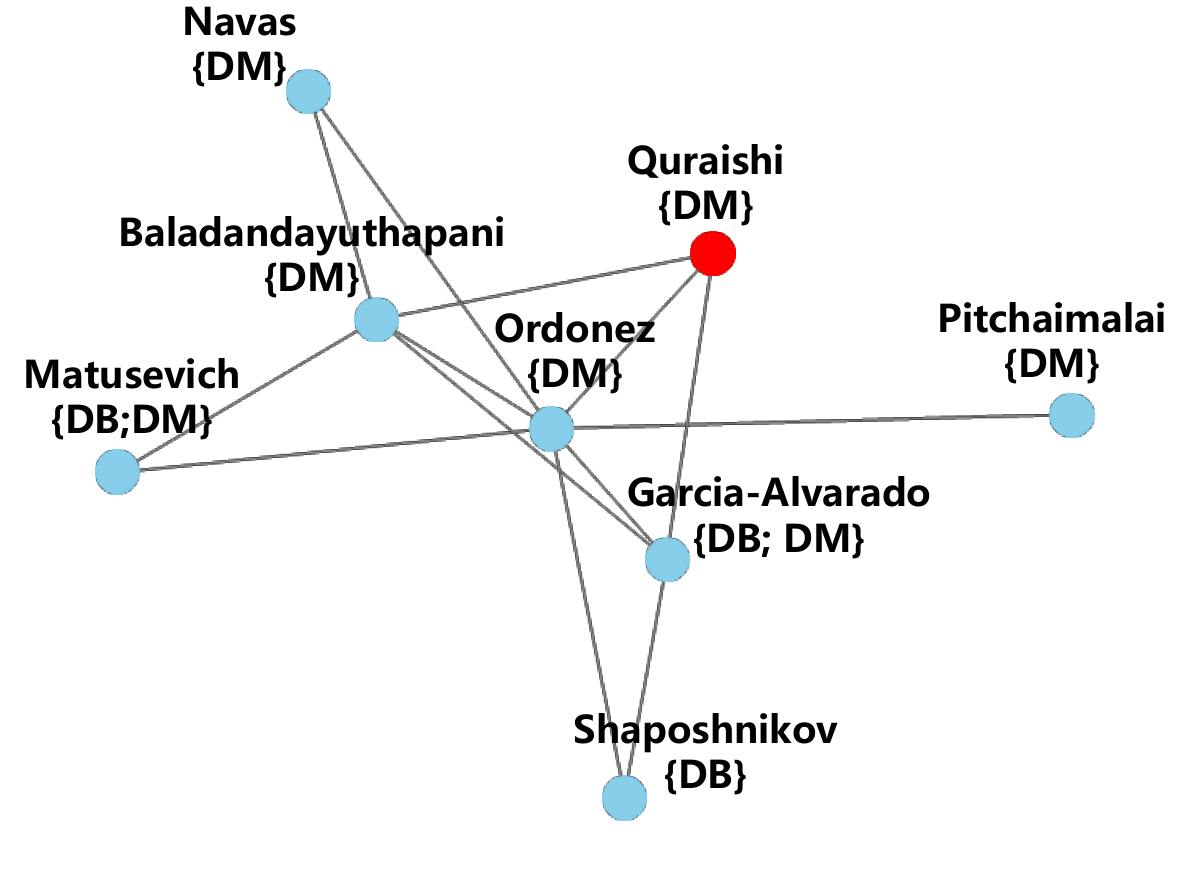}
        \caption{(c) LSADEN}
        \label{fig:case_03}
    \end{subfigure}
        \begin{subfigure}[b]{0.49\columnwidth}
        \includegraphics[width=1\linewidth]{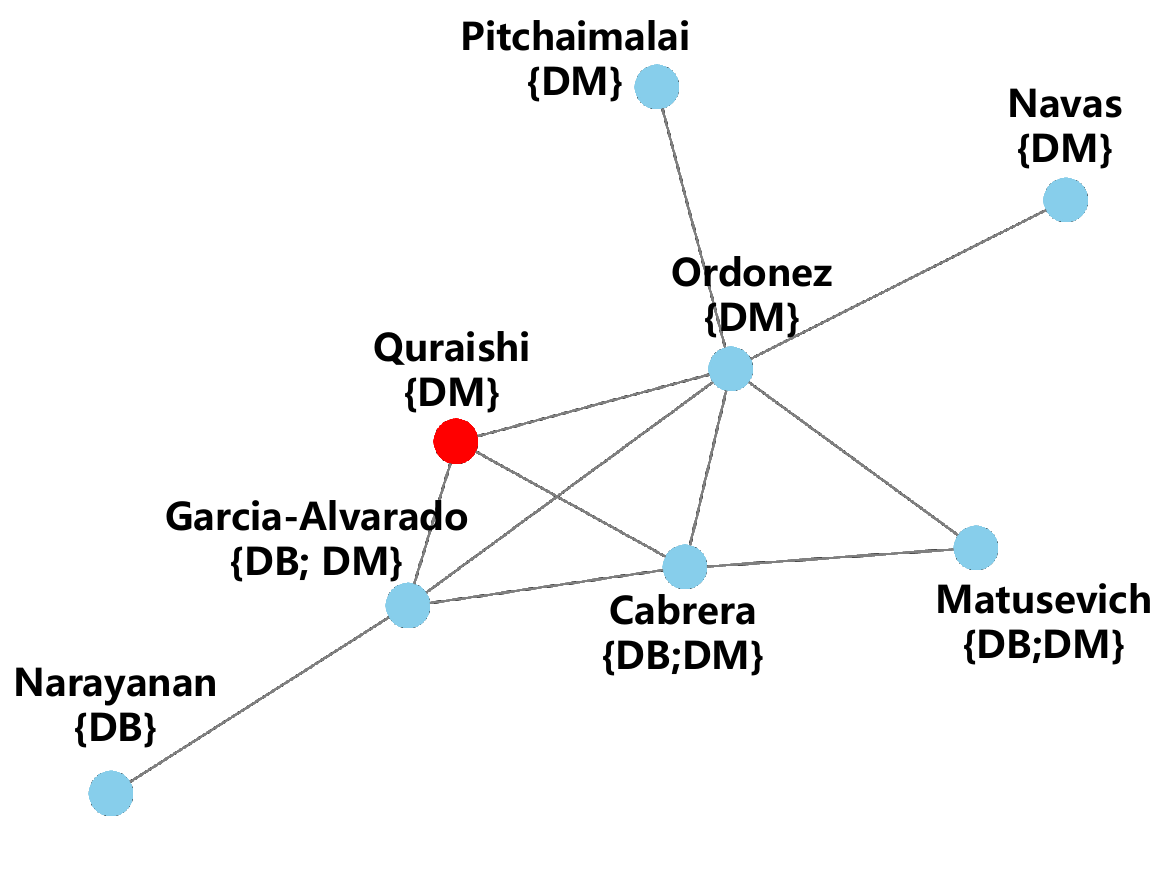}
        \caption{(c) SLDRG}
        \label{fig:case_04}
    \end{subfigure}

    \caption{Results of case study.
  The nodes represent scholars, labeled by their last names (first names are omitted for clarity). Due to the large number of community nodes identified by ACQ, last names are also omitted here.}
    \label{figure:case study}
\end{figure}

We conduct a case study on the Aminer+CA dataset, where Aminer\footnote{https://cn.aminer.org/data} is a network of scientific collaborators from DM, DB, and ML fields. It consists of 25,165 nodes and 42,358 edges, with author locations normalized to map onto the road network of California\textsuperscript{\ref{california}}, denoted as “CA”. Assume author “Quraishi” plans to invite scholars for a DM seminar. “DM” is the query keyword, and Quraishi’s institution location is the query location.
Figure \ref{figure:case study} shows the results. The community identified by GSSAC specializes in the DM field and has the shortest average travel time to the query location. The institutions are mainly located in the U.S. and Mexico, and the community’s $t_{coe}$ is 0.1039. In contrast, ACQ identifies a large community of nearly 300 nodes, with a $t_{coe}$ value of 0.0125, much lower than GSSAC’s. LSADEN and SLDRG find communities with closely located members but include members outside the DM field, such as “Shaposhnikov” and “Narayanan”. GSSAC identifies a 3-core, ensuring members have prior collaboration experience. In summary, scholars identified by GSSAC are geographically near Quraishi's institution, share the same research domain, and have prior collaboration experience, increasing their likelihood of being invited to the seminar.

\section{Related Work}\label{Related Work}
This section provides a brief review of related work on community search in social networks, attributed social networks, and road-social networks.

\subsection{Community search in social networks. } 
It aims to identify communities containing specific query nodes \cite{Fang_02, Wang_02}, also known as local community detection \cite{Feng_01, Ni_03, Ni_04, Guo_02}. Dense subgraph models for community search include $k$-core \cite{Sozio_01, Cui_01}, $k$-truss \cite{Huang_02, Xie_01}, $k$-clique \cite{Gergely_01},  and local modularity \cite{Wenjiang_01}. The $k$-core  can be mined using core decomposition \cite{Batagelj_01} or greedy algorithms \cite{Sozio_01}. Cui et al. \cite{Cui_01}  explored $k$-core with a local expansion approach, while Huang et al. \cite{Huang_02} focused on $k$-truss with an index-based method for dynamic graphs. Unlike these studies focusing only on network structure, our work additionally incorporates keyword and location information.

\subsection{Community search in attributed social networks. }
Real-world networks often include keyword and location information \cite{Fang_02}. Keyword-based studies typically identify communities based on query nodes and keywords \cite{Fang_01}. Fang et al. \cite{Fang_01} defined community-shared keywords to find communities with the most shared keywords. Huang et al. \cite{Huang_02} proposed a keyword score function to assess semantic cohesiveness by calculating the average frequency of query keywords within communities. Ye et al. \cite{Ye_01} considered both semantic similarity between users and query keywords and similarity among users within communities to ensure semantic cohesiveness.

Location-based studies aim to find communities based on network topology and locations \cite{Fang_03, Sachan_01, Liang_01}. Fang et al. \cite{Fang_03}  ensure spatial cohesiveness using the minimum covering circle. Chen et al. \cite{Chen_01} identifies communities with average member distances below a threshold. Ni et al. \cite{Ni_01, Ni_02, Ni_05} minimized the average distance between community members. Some studies combine keyword and location information, i.e., the ACOC search problem \cite{Luo_01}, which ensures communities contain all query keywords while minimizing the community diameter. However, these studies either ignore location attributes or use Euclidean distance, while real users travel along road networks. Unlike them, our work considers both keywords and locations, measuring distances along real road networks.

\subsection{Community search in road-social networks. }
Community search in road-social networks can help organize instantaneous activities \cite{Guo_01}. Guo et al. \cite{Guo_01} searched for maximal $k\text{-}core$ communities containing query nodes and measuring spatial cohesiveness with actual travel distances. Li et al. \cite{Qiyuan_01} used $k\text{-}core$ for structural cohesiveness and minimizing travel distances for spatial cohesiveness. These studies \cite{Guo_01, Qiyuan_01} focus on network structure and road locations but ignore node keywords. Al-Baghdadi et al. \cite{Al_01} required detected communities to contain at least one query keyword and limit the average road distance between nodes. However, it overlooks the overall keyword matching and the dynamic nature of travel times due to factors like traffic. Unlike these studies, our work integrates both keyword and location information while considering dynamic travel time changes on real road networks.
\section{Conclusion}\label{Conclusion}
In this paper, we study the problem of identifying semantic-spatial aware $k$-core.  
To address this problem, we propose two local methods, ESSAC and GSSAC. 
Both methods start from the query node and gradually expand outward to detect the semantic-spatial aware $k$-core.
ESSAC explores many $k$-cores with poor semantic and time-dependent spatial cohesiveness that do not require further expansion. To improve efficiency, GSSAC expands a single $k$-core by optimizing both semantic and time-dependent spatial cohesiveness.
To calculate the semantic similarity between two keywords, we design prompts to process the keywords with GPT-3.5-Turbo and obtain their embeddings using text-embedding-3-small.
Experimental results show that GSSAC performs better than baselines in terms of community quality.

\section*{Acknowledgments}
This work was supported by the National Natural Science Foundation of China [No.62106004, No.62272001, and No.62206004]. 

\bibliographystyle{IEEEtran}
\bibliography{references}

\vfill

\end{document}